\documentclass[referee]{aa} 

\usepackage{graphicx}
\usepackage{natbib}
\usepackage{txfonts}
\begin{document}

   \title{ Abundance ratios of red giants in low mass ultra faint dwarf spheroidal galaxies}

\titlerunning{ Abundances in UfDSph red giants}
   \author{P. Fran\c cois
          \inst{1,2}
          \and
          L. Monaco\inst{3}
            \and
          P. Bonifacio \inst{1}    
          \and
                    C. Moni Bidin \inst{4} 
          \and
                    D. Geisler \inst{5}
           \and
                   L. Sbordone \inst{6}\thanks{Based on observations made with ESO Telescopes at the La Silla Paranal Observatory under programme ID 085.B-0367(A)}
          }
   \institute{ GEPI, Observatoire de Paris, PSL Research University, CNRS, Univ Paris Diderot, Sorbonne Paris Cité, 61 Avenue de l'Observatoire, 75014 Paris, France \\
              \email{patrick.francois@obspm.fr} 
         \and
             Universit\' e de Picardie Jules Verne, 33 rue St Leu, Amiens, France 
             \and 
             Departamento de Ciencias Fisicas, Universidad Andres Bello, Republica 220, Santiago, Chile
             \and
             Instituto de Astronom\'ia, Universidad Cat\'olica del Norte, Av. Angamos 0610, Antofagasta, Chile            
             \and
             Department of Astronomy Faculty of Physical and Mathematical Sciences, University of Concepci\' on, Chile
             \and
             Millennium Institute of Astrophysics, Pontificia Universidad Cat\'olica de Chile, Vicu\~na Mackenna 4860, Macul Santiago, Chile
             }

   \date{Received ; accepted }

 
  \abstract
   { Low mass dwarf spheroidal galaxies are key objects for our understanding of the chemical evolution of the 
pristine Universe and the Local Group of galaxies.  Abundance ratios in stars of these objects 
can be used to better understand their star formation and  chemical evolution. }
   { We report on the  analysis of  a sample of  11 stars belonging to 5 different 
ultra faint dwarf spheroidal galaxies (UfDSph) based on X-Shooter spectra obtained at the VLT. }
   {  Medium resolution spectra have been used to determine the detailed chemical composition of their atmosphere. 
    We performed a standard 1D LTE analysis to compute the abundances.}
   {
Considering all the stars as representative of the same population of low mass galaxies, we found 
that the [$\alpha$/Fe] ratios  vs [Fe/H]  decreases as the metallicity of the star
increases  in a way similar to what is found for the population  of stars belonging to dwarf 
spheroidal galaxies.  The main difference is that the  solar [$\alpha$/Fe] is reached at a much lower 
metallicity for the UfDSph than the dwarf spheroidal galaxies.  

We report for the first time the abundance of strontium in CVnI.  The star we analyzed in this galaxy has a very high [Sr/Fe] and a very low 
upper limit of barium which makes it a star with an exceptionally high [Sr/Ba] ratio. 

Our results  seem to indicate that the galaxies which have produced the bulk of their stars before the reionization  (fossil galaxies) have lower $[X/Fe]$  ratios  at a given metallicity 
than the galaxies that have experienced a discontinuity  in their star formation rate (quenching).
 
    }
   {}


   \maketitle
%

\section{Introduction}

$\Lambda$ cold dark matter cosmological models are in agreement with many observable phenomena, but some discrepancies are 
found on small scales.  In particular, this model predicts too many  dark-matter sub-halos (a factor of 50)  that the number of observed  dwarf galaxies  \citep{moore1999}. A solution to this problem was put forward by \citet{bullock2000}  who suggested that gas accretion in low-mass halos could be suppressed by the photo-ionization mechanism during the reionization of the Universe.  The observed dwarf satellites correspond  to the small fraction 
of halos that accreted enough amounts of gas before reionization.  Based on this hypothesis,  \citet{ricotti2005} proposed that dwarf galaxies could be 
classified in three different classes depending on the occurrence of their star formation relatively to the reionization event.  Dwarf galaxies that formed most of their stars prior to the reionization are  classified as "true fossils".
From this classification,it appears that  some stars we observe
today in the so-called ultra faint dwarf spheroidal galaxies (UfDSph, \citet{belo2007})  could be "survivors" of the reionization period. \citet{brown2014}  analyzed the star formation history of six UfDSphs  (Bootes I, Canes Venatici II, Coma Berenices, Hercules, Leo IV and Ursa Major I). They concluded that five out them are best fit by a star formation history  where at least 75 $ \%$  of the stars formed by z $\simeq$ 10  and 100  $ \%$ of the stars
formed by z $ \simeq$ 3 i.e. 11.6~Gyrs ago, supporting  the hypothesis of a quenching of the star formation   by a global external influence such as reionization.  
The detailed chemical composition of stars in ultra-faint dwarf spheroidal galaxies  is therefore a important tool to probe the early evolution of the local group. 
This paper reports on the detailed abundance determination of stars belonging to the UfDSphs Bootes II, 
Canes Venatici I, Canes Venatici II, Hercules and  Leo IV.  Among these fives galaxies, two  (Leo IV and Hercules) are consistent with the hypothesis 
that the bulk of their stars were formed  before reionization  \citep{weisz2014}.   However, their conclusion has been challenged \citep{brown2014} . Three of our galaxies were analyzed by \citet{brown2014}. Using deep and high S/N ACS imaging over a wide field, they  found that these galaxies (and three others, among them BooI) were consistent with the hypothesis that reionization  ended star formation in all of them. 

\citet{webster2015}  have modeled the chemical evolution of the six UfDsph galaxies studied by \citet{brown2014} , among them Boo I, CVnI, Hercules and LeoIV). They showed that two single-age bursts cannot explain the observed [$\alpha$/Fe] verus [Fe/H]  distribution in these galaxies.  They suggested an alternative scenario in which star formation is continuous except for short interrruptions.    
From their table 1,  Hercules and LeoIV have the same likelihood to have quenched or non-quenched star formation history. 
If we take into account of the studies made by \citet{weisz2014} and \citet{brown2014},   these two galaxies can be considered  as "fossil" galaxies.
  On the other hand,  Boo I and CVnII have a higher likelihood to have an extended star formation history. 
Based on the results from \citet{webster2015} we  classify  Her and LeoIV as "fossil  galaxies" and CVnII, BooI and CVnI as galaxies with extended star formation history.

Before going into the details of our analysis, we would like to remind the most important characteristics of each galaxy for which we obtained mid-resolution spectra with the ESO-VLT and the X-Shooter spectrograph.


\subsection { Bootes II} 
The discovery of Bootes II was reported by \citet{walsh2007} as an over density on the Sloan Digital Sky Data	 Release 5 ( hereafter SDSS DR5) distribution. From isochrone fitting techniques and accurate color-magnitude diagram, they described Bootes II as an old (12~Gr), metal-poor ([Fe/H] $\simeq$  $-$2.00  dex) galaxy  with a distance estimated
at 60 kpc.  MMT/Megacam imaging  in Sloan g and r  \citep{walsh2008}  led a to a revised distance of 42 $\pm 2$ kpc.
From follow-up spectroscopy of five member stars, \citet{koch2008} found a mean metallicity of [Fe/H] = $-$1.79 $\pm 0.05 $ dex. 
 This determination relies on an old calibration of the Ca triplet which was revised later on.
 \cite{koch2014}  made a detailed chemical analysis of the brightest confirmed member  star in Boo II using Keck/Hires and derived a very low
metallicity of [Fe/H] = $-$2.93 dex using an updated Ca triplet calibration.  They also found a high $[\alpha/Fe]$ ratio compatible with the $\alpha$-enhanced plateau value of the galactic halo. 

\subsection { Canes Venatici I} 

Canes Venatici I  was discovered in 2006 by \citet{zucker2006}
as a stellar over density in the north Galactic cap using the Sloan Digital Sky Survey Data Release 5.
From the tip of the red giant branch, they concluded that the Galaxy was at a distance of $\simeq$ 220 kpc. 
The first deep color-magnitude diagrams of the Canes Venatici I (CVn I) dwarf galaxy were provided by \cite{martin2008}  from observations 
with the wide-field Large Binocular Camera on the Large Binocular Telescope. Interestingly, their analysis revealed a dichotomy in the stellar populations of CVn I which harbors an old ($ \ge$ 10~Gyrs), metal-poor ([Fe/H] $\simeq$ $-$2.0),  and spatially extended population along with a
much younger, more metal-rich and spatially more concentrated population.  However, the claim of a young population in   Canes Venatici I
has not been supported by more recent studies  \citep{ural2010, okamoto2012} .
 \cite{martin2008} derived a distance modulus of $(m - M)_0 = 21.69 \pm 0.10 $
or $D= 218 \pm 10$ kpc . 
\citet{okamoto2012} confirmed the distance modulus using deep images taken with the Subaru/Suprime-Cam imager obtaining a  distance modulus of  $ (m - M)_0 = 21.68 \pm$ 0.08  (216 $\pm$ 8 kpc) . 
\citet{kirby2010}  determine the  abundances of Fe  and several $\alpha$ elements in a sample of 171 stars using medium resolution  spectra 
 (R $\sim$ 7000) obtained with Keck/DEIMOS  and found metallicities ranging from $-$1.0 dex to $-$3.3 dex.  No high resolution spectroscopy has been performed so far. 

\subsection { Canes Venatici II } 

The UfDSph  Canes Venatici  II is one  of the four UfDSph discovered by \citet{belo2007} in the Sloan digital Sky Survey.  Follow-up spectroscopic observations were performed in 2008 by \citet{kirby2008} who analyzed 16 stars.  
They used DEIMOS on the Keck II telescope to obtain spectra at R $\simeq $ 6000 over a spectra range of roughy 6500-9000 \AA.
They derived a  mean metallicity of [Fe/H] = $-$2.19 $\pm$ 0.05 dex with a dispersion of 0.58 dex.  \citet{vargas2013} computed the $[\alpha / Fe]$ ratios in 8 stars of this galaxy and found an increase of the $[\alpha / Fe]$ as  metallicity decreases with a solar ratio  at [Fe/H] $\simeq $ $-$1.30 dex to reach on average  an $[\alpha / Fe] \simeq  0.5$ dex at  [Fe/H] $\simeq $ $-$2.50 dex. The distribution of [Ca/Fe] and [Ti/Fe]  abundance ratios tends to point towards the presence of a significant scatter at low [Fe/H].   The metallicity was later revised by \cite{vargas2013} who found [Fe/H] = $-$2.18 $\pm$ 0.06 dex.

\subsection { Hercules} 

Hercules is a dwarf spheroidal satellite of the Milky Way, found at a distance of 138 kpc.

This UfDSph has been discovered by \citet{belo2007}. 
\citet{cole2007} performed deep wide-field photometry in B, V and r of this galaxy using the Large Binocular Telescope   down to 1.5 mag below the main sequence turn-off and found that  the Hercules dwarf is highly elongated suggesting tidal disruption as a likely cause. 
\citet{simon2007} obtained a  first estimate of the metallicity using Keck-DEIMOS spectroscopy of 30 stars finding  [Fe/H] $\simeq$ $-$2.27   with a dispersion of 0.31 dex. 
\citet{koch2008} analyzed 2 red giants and derived a metallicity of [Fe/H] $\simeq$ $-$2.00 dex with strong enhancements in Mg and O and a high deficiency in the neutron capture elements.
Later, \citet{aden2009} analyzed 11 stars in Hercules  and obtained a metallicity spread ranging from [Fe/H] = $-$2.03 to $-$3.17 dex. They also  found that the red giant branch stars  are more metal-poor than previously estimated by photometry. A comparison of their spectroscopic stellar parameters with isochrones indicates that the giants in Hercules are older than 10 Gyrs. 
\cite{koch2013} analyzed a new sample of four red giants  and confirmed the high level of depletion of the neutron capture elements and suggested 
that the chemical evolution of Her was dominated by very massive stars. 
Deep g,i-band DECam stellar photometry of the Hercules Milky Way satellite galaxy, and its surrounding field, out to a radial distance of 5.4 times the tidal radius was done by \cite{rode2015}.They identified nine extended stellar substructures associated with the dwarf, preferentially distributed along the major axis of the galaxy demonstrating that Hercules is a strongly tidally disrupted system. 

\subsection { Leo IV} 

The dwarf galaxy Leo IV was discovered by \citet{belo2007} along with Coma Berenices, Canes Venatici II and Hercules.  
\cite{simon2007}  obtained Keck/DEIMOS spectra of a sample of stars belonging to this galaxy and derived a metallicity of [Fe/H] = $-$2.31 $\pm$ 0.10 dex.
Adopting a reddening E(B - V) = 0.04 $\pm$ 0.01 mag and a metallicity of [Fe/H] = $-$2.31 $ \pm $ 0.10 dex.
 \citet{moretti2009} derived a distance of 154 $\pm$ 5 kpc.   The first determination of the chemical composition of stars in Leo IV was done by \cite{simon2010}. They  obtained high resolution Magellan/MIKE spectra of the brightest star in Leo IV and measured an iron abundance [Fe/H] = $-$3.2 dex with an $\alpha$ element enhancement similar to what is found in the milky way halo.  Interestingly, this star is among the most metal poor stars found in UfDSphs. 
 \citet{okamoto2012}  estimated the average age of the stellar population to be 13.7 Gyrs by overlaying Padova isochrones.  
We present in this paper, the determination of the detailed chemical composition of two stars belonging to LeoIV.
{ \citet{vargas2013} revised the metallicity and found [Fe/H] = $-$2.89 $ \pm $ 0.11 dex .}

\section{Observations}

 The aim of these observations was the study of the metal-poor population of stars belonging to UfDSphs.
Therefore the sample is biased towards the brightest targets among the metal poor sample of these galaxies.
Target stars were selected from among the most metal-poor radial velocity member stars with V $<$ 20.0 in each galaxy, and
were extracted from published low-resolution studies \citep{kirby2008, koch2009}. All
of the targets have putative metallicities [Fe/H] $<$- 2.0 dex , 9 out of 11 having [Fe/H]<-2.6, i.e. more metal-poor
than any Galactic globular cluster. From their CaT index, seven stars have [Fe/H] $<$ -3.0 dex and are, therefore, extremely metal-poor.

The observations were performed in service mode with the ESO-Kueyen  telescope (VLT UT2) 
and the high-efficiency spectrograph X-Shooter \citep{dodo2006, vernet2011}. 
The list of targets is given in table \ref{star_id}.
  \begin{table*}
      \caption[]{Target coordinates, Signal to noise ratios  and radial velocities}
  \begin{tabular}{l l l c c c c }
            \hline
            \noalign{\smallskip}
Galaxy &    Object                               &        Other ID           &     RA                     & DEC                  & SNR at 500 nm  &  Vr \\
            &                                              &                                &                       &                 &        &       km~s$^{-1}$              \\
Boo~II &  SDSS J135801.42+125105.0 & Boo II -- 7    & 13h58m01.0s  & 12d51m04.7s  &    55   &  -138  \\
Boo~II &  SDSS J135751.18+125136.9 & Boo II -- 15  &13h57m51.2s   & 12d51m36.6s    &   60   & -119\\
 Leo IV &  SDSS J113255.99-003027.8 & Leo IV -- S1     & 11h32m56.0s   & -00d30m27.8s & 35 & 126 \\
Leo IV& SDSS J113258.70-003449.9 &       & 11h32m58.7s   & -00d34m50.0s  &   50   &  129\\
CVn II  & SDSS J125713.63+341846.9 &     & 12h57m13.6s  &  +34d18m46.9s &  25 &  135\\
CVn I& SDSS J132755.65+333324.5 &      & 13h27m55.7s  & +33d33m24.5s  &  30  &  21\\
CVn I& SDSS J132844.25+333411.8 &       & 13h28m44.3s  & +33d34m11.8s  &  35 &  21\\
Her & SDSS J163044.49+124947.8 &      & 16h30m44.5s   & +12d49m47.9s  &    45  &  35 \\
Her & SDSS J163059.32+124725.6 &       & 16h30m59.3s   & +12d47m25.6s  &   50 & 36  \\
Her & SDSS J163114.06+124526.6 &        & 16h31m14.1s   & +12d45m26.6s &  70 &  49\\
 Her & SDSS J163104.50+124614.4 &     & 16h31m04.5s   & +12d46m14.5s  & 35  & 40\\
              \noalign{\smallskip}
            \hline
         \end{tabular}
         \label{star_id}\\
   \end{table*}

The observations have been performed in staring mode with 1x1 binning and the integral field unit (IFU).
We used the IFU as a slicer with three 0.6$\arcsec$ slices. This corresponds to 
a resolving power of R = 7900 in the ultra-violet arm (UVB) and R = 12600 in the visible arm (VIS). The stellar light is divided in three arms by X-Shooter; 
we analyzed here only the UVB and VIS spectra. The stars we observed are very faint and have most of their flux in the blue part of the spectrum, so that 
the signal-to-noise ratio (S/N) of the infra-red spectra is too low to allow a reliable chemical abundance analysis. Moreover, the sky contamination in staring mode affects strongly the  stellar 
spectrum. The spectra were reduced using the X-Shooter pipeline 
\citep{goldo2006}, which performs the bias and background subtraction, cosmic-ray-hit removal \citep{vandok2001}, sky subtraction 
\citep{kelson2003}, flat-fielding, order extraction, and merging. However, the spectra were not reduced using the IFU pipeline recipes. 
Each of the three slices of the spectra were instead reduced separately in slit mode with a manual localization of the source and the sky. 
This method allowed us to perform the best possible extraction of the spectra, leading to an efficient cleaning of the remaining cosmic ray hits,
 but also to a noticeable improvement in the S/N, thanks to the optimal extraction 
pipeline routine of X-Shooter. Using the IFU can cause some problems with the sky subtraction because there is only $\pm$1$\arcsec$ on both sides 
of the object. In the case of a large gradient in the spectral flux (caused by emission lines), the modeling of the sky-background signal can be of
 poor quality owing to the small number of points used in the modeling.  As we made our analysis only in  the UVB and VIS spectra of X-Shooter, 
 only few lines are affected by this problem.  
 
 We used the strong lines of magnesium to determine the radial velocities of the stars using the cross-correlation of the synthetic spectrum with the observed spectrum. Heliocentric corrections 
 have been also applied.  The radial velocities of the stars are reported in table \ref{star_id}.  Typical errors of 5 km~s$^{-1}$ have been estimated by computing the dispersion of the measurements of the radial velocities  on the individual spectra before combining them  for the abundance determination.  
 The results are in good agreement with the systemic radial velocities of the parent galaxies.

   \begin{table*}
      \caption[]{Log of the observations. All the exposures are of 2950 seconds. }
  \begin{tabular}{l l c }
            \hline
            \noalign{\smallskip}
                OBJECT             &          	TIMESTAMP                  	  	\\
	 SDSS J113258.70-003449.9 / Leo IV &   	2011-02-07T07:11:20.295	  	 \\
	 SDSS J113258.70-003449.9 / Leo IV   & 	2011-02-07T08:17:40.984	 	 \\
	SDSS J125713.63+341846.9 / CVn II    &	2010-04-02T06:36:39.302	  	 \\		
	SDSS J125713.63+341846.9 / CVn II    &	2011-04-09T04:33:13.527	   \\
	SDSS J125713.63+341846.9 / CVn II    &	2011-04-09T05:43:44.497	 	 \\
	SDSS J132844.25+333411.8 / CVn I      &	2011-05-04T13:01:06    	 	 \\		
	SDSS J132844.25+333411.8 / CVn I      &	2011-04-27T04:23:11.240   	 \\	
	SDSS J113258.70-003449.9 / Leo IV   &	2010-05-10T03:02:20.251	  	 \\
	SDSS J113258.70-003449.9 / Leo IV   &	2010-05-10T04:10:07.978   \\	
	SDSS J135751.18+125136.9 / Boo II -- 15  &	2010-05-12T02:26:01.360	  	 \\
	SDSS J135751.18+125136.9 / Boo II -- 15  &	2010-05-12T03:21:50.970    	 \\	
	SDSS J113255.99-003027.8 / Leo IV -- S1   &	2010-06-10T02:12:37.200	    	 \\
	SDSS J135801.42+125105.0 / Boo II -- 7   &	2010-06-10T03:35:48.990	 	 \\
	SDSS J163044.49+124947.8 / Her      &	2011-05-13T14:39:48    	 	 \\		
	SDSS J132755.65+333324.5 / CVn I      &	2010-06-12T02:13:14.498	  	 \\	
	SDSS J163044.49+124947.8 / Her      &	2010-06-12T05:06:48.771	  	 \\
	SDSS J163044.49+124947.8 / Her      &	2010-07-09T05:11:07.059	 	 \\	
	SDSS J163059.32+124725.6 / Her      &	2010-07-10T04:11:23.800	   	 \\
	SDSS J163104.50+124614.4 / Her      &	2010-07-13T04:04:06.392	   	 \\
	SDSS J135801.42+125105.0 / Boo II -- 7   &	2010-08-03T00:28:32.920	   	 \\
	SDSS J135801.42+125105.0 / Boo II -- 7   &	2010-08-04T00:45:15.158	   	 \\
	SDSS J163114.06+124526.6 / Her      &	2010-08-03T02:43:58.485	   	 \\
	SDSS J135801.42+125105.0 / Boo II -- 7   &	2010-08-05T00:27:02.020	  	 \\
	SDSS J163104.50+124614.4 / Her      &	2010-08-05T01:34:40.556    	  \\
	SDSS J163104.50+124614.4 / Her     &	2010-08-09T03:03:43.065	   	 \\
	SDSS J163104.50+124614.4 / Her      &	2010-08-10T01:43:39.930	 	 \\
            \noalign{\smallskip}
            \hline
         \end{tabular}
         \label{obs_log}\\
   \end{table*}

 \section {Analysis}

The effective temperature was derived from the $(g - i)$ colors   \citep{koch2009} 
for the two Bootes stars  and the V and $I_\mathrm{C}$ colors 
taken from \citet{kirby2008} for the remaining stars.  $(g - i)$ have been transformed into $(V-I_\mathrm{C})$ using the relation given by \citet{jordi2006}.
The reddening correction is from \citet{schlegel1998}. 
We adopted  the calibration of \citet{ramirez2005}, use of the \citet{alonso1999}
calibration would result in temperatures
that are  100  K to 150 K hotter. 
Note that all the published color calibrations are ill defined
for very metal-poor giants, due to a scarcity of calibrators. 
The  \citet{ramirez2005} sample of calibrators has more
metal-poor giants than the  \citet{alonso1999}
sample, hence our choice.
The surface gravities have been obtained from the photometry and calculated
using  the  standard  relation  between  $log g$,  mass,  $T_\mathrm{eff}$ and  $M_\mathrm{bol}$  relative  to  the  Sun,
assuming the solar values $T_\mathrm{eff,\sun}=5790~K$, $log g = 4.44$ and $M_\mathrm{bol}= 4.72$. We assumed also 0.8 $M_{\odot}$
for the mass of the giant stars which have been observed.  Distance moduli have been taken from \citet{walsh2008} for BooII, from \citet{kuehn2008} for  CVnI , from \citet{greco2008} for  CVnII, from \citet{musella2012}  for Her and from \citet{moretti2009} for LeoIV.

We carried out a classical 1D LTE analysis using 
OSMARCS model atmospheres  \citep{gustafsson1975, plez1992, edvardsson1993, asplund1997, gustafsson2003}.
The abundances used in the model atmospheres
were solar-scaled with respect to the \citet{grevesse2000} solar abundances,
except for the $\alpha$ elements that are enhanced by 0.4\,dex. We corrected the resulting abundances by taking into account the difference
between  \citet{grevesse2000}  and \citet{caffau2011b}, \citet{lodders2009}  solar abundances. 

The abundance analysis was performed using the LTE spectral line analysis code turbospectrum (Alvarez \& Plez 1998; Plez 2012), that treats scattering in detail. The carbon abundance was determined by fitting the CH band  near at 430 nm (G band). The molecular data corresponding to the CH band are described in \citet{hill2002} and Plez et al. (2008).

Two stars (HD~165195 and HE1249-3121) with published detailed abundances \citep{gratton1994, allen2012} obtained using high resolution spectra have been used to check the validity of our abundance determinations.  For these two stars, we retrieved X-shooter spectra and recomputed the abundances of the elements measured in our sample of UDSph galaxies stars.  Our results are in agreement within 0.1 dex with the published abundances.

The adopted stellar parameters can be found in  Table \ref{star_param}.

   \begin{table*}
      \caption[]{Stellar parameters}
  \begin{tabular}{l c c c c}
            \hline
            \noalign{\smallskip}
  Star               &   $T_\mathrm{eff}$ & logg & $\xi$               & [Fe/H] \\
                        & K                            & dex & km~s$^{-1}$ & dex \\
SDSS J163044.49+124947.8 / Her &   4700  & 1.40  & 2.1 & -2.54 \\
SDSS J163059.32+124725.6 / Her   &  4600  & 1.20  & 2.2 & -2.85\\
SDSS J163104.50+124614.4 / Her   & 4870  & 1.70  & 2.2  & -2.55\\
SDSS J163114.06+124526.6 / Her   & 4750  & 1.40  & 2.0  & -2.30\\
SDSS J113255.99-003027.8 / Leo IV -- S1    & 4500  & 1.10  & 2.5  & -2.90\\
SDSS J113258.70-003449.9 / Leo IV    & 4800  & 1.50  & 2.4  & -2.20\\
SDSS J125713.63+341846.9 / CVn II    & 4590  & 1.20  & 2.0  & -2.60\\
SDSS J135801.42+125105.0 / Boo II -- 7    & 4910  & 2.50  & 2.0  & -3.10\\
SDSS J135751.18+125136.9 / Boo II -- 15   & 4980  & 2.60  & 2.3  & -3.00\\
SDSS J132844.25+333411.8 / CVn I   & 4450  & 0.81  & 2.0  & -2.50\\
SDSS J132755.65+333324.5 / CVn I    & 4350  & 0.72 &  2.3  & -2.20 \\
            \noalign{\smallskip}
            \hline
         \end{tabular}
         \label{star_param}\\
   \end{table*}

    \begin{table}
      \caption[]{List of  lines used for the analysis. Hfs data for barium  are from McWilliam \& Preston (1995). }
  \begin{tabular}{l c c c}
   \hline
   \noalign{\smallskip}
Wavelength &  Ion & $\chi_{exc}$ & log gf \\
   \hline
      \noalign{\smallskip}

5889.950  & Na~I  & 0.00    &   0.11     \\      
5895.924 &  Na~I   & 0.00    &  -0.19    \\       
4351.921  & Mg~I  & 4.34    &  -0.53     \\         
4571.102  & Mg~I   & 0.00    &  -5.39     \\          
5172.698  & Mg~I   & 2.71    &  -0.38     \\          
5183.619  & Mg~I   & 2.72    &  -0.16     \\          
5528.418  & Mg~I   & 4.34   &   -0.34      \\     
3944.016  & Al~I   & 0.00   &   -0.64        \\       
4226.740  & Ca~I   & 0.00   &    0.24       \\                    
4283.014  & Ca~I   & 1.89    &  -0.22       \\          
6122.226  & Ca~I  & 1.89   &   -0.32        \\       
6162.173  & Ca~I   & 1.90   &   -0.09       \\        
4318.659 &  Ca~I   & 1.90   &   -0.21       \\       
4030.763 & Mn~I  & 0.00 &    -0.48        \\       
4033.072 & Mn~I  & 0.00  &   -0.62       \\            
3920.269  & Fe~I & 0.12   & -1.75   \\            
 3922.923 &  Fe 1 & 0.05   &  -1.65    \\          
 4045.825 & Fe~I  & 1.48   &   0.28    \\          
 4063.605 & Fe~I  & 1.56   &   0.07     \\         
 4071.749 & Fe~I  & 1.61   &  -0.02  \\            
 4143.878 & Fe~I  & 1.56   &  -0.46  \\            
 4181.764 & Fe~I  & 2.83   &  -0.18  \\            
 4191.437 & Fe~I  & 2.47   &  -0.73   \\           
 4202.040 & Fe~I  & 1.48  &   -0.70  \\            
 4260.486 & Fe~I  & 2.40   &  -0.02   \\           
 4282.412 & Fe~I  & 2.17  &   -0.82   \\           
 4307.912 & Fe~I  & 1.56   &  -0.07   \\           
 4383.557 & Fe~I  & 1.48   &   0.20    \\          
 4404.761 & Fe~I   &1.56   &  -0.14    \\          
 4415.135 & Fe~I  & 1.61   &  -0.61    \\          
 4427.317 & Fe~I   & 0.05   &  -2.92     \\         
 4459.100 & Fe~I  & 2.18   &  -1.28    \\          
 4461.660 & Fe~I  & 0.09   &  -3.20    \\          
 4489.748 & Fe~I  & 0.12   &  -3.97    \\          
 4494.573 & Fe~I  & 2.20   &  -1.14     \\         
 4531.158 & Fe~I  & 1.48   &  -2.15   \\           
 4920.514 & Fe~I  & 2.83   &   0.07    \\          
 5083.345 & Fe~I  & 0.96   &  -2.96     \\         
 5194.949 & Fe~I  & 1.56   &  -2.09   \\           
 5371.501 & Fe~I  & 0.96   &  -1.65   \\           
 5405.785 & Fe~I  & 0.99   &  -1.84    \\          
 5429.706 & Fe~I  & 0.96   &  -1.88    \\          
 5446.924 & Fe~I  & 0.99   &  -1.91   \\           
 5455.624 & Fe~I  & 1.01 &    -2.09    \\    
 6136.615 &  Fe~I  & 2.45    &  -1.40   \\    
 6191.571 &  Fe~I  &  2.43   &   -1.42   \\    
 4118.782  & Co~I  &  1.05   &   -0.49   \\    
 4121.325  & Co~I  &  0.92   &   -0.32    \\   
 5476.921  & Ni~I  &  1.83   &   -0.89     \\   
 4077.724  & Sr~II  &  0.00    &   0.167     \\  
 4215.520  & Sr~II  &  0.00  &    -0.145    \\ 
  4934.095  & Ba~II  &  0.00   &   hfs   \\
 5853.688  & Ba~II  &  0.60   &    hfs    \\  
  6141.727  & Ba~II  &  0.70 &     hfs    \\

            \noalign{\smallskip}
            \hline
         \end{tabular}
         \label{linelist}\\
   \end{table}

 We measured the equivalent widths for  a list of \ion{Fe}{i} lines given in Table \ref{linelist}.   
 With the assumed stellar parameters, we first checked the micro turbulent velocity using the method of the curves of growth (see for example
 \citet{lemasle2008} 
 
 We checked the excitation temperature  and   refined  the determination of the micro turbulent velocity parameters 
 using the standard trends  abundance vs excitation temperature  and abundance vs equivalent width.  
The iron excitation is satisfied for our adopted temperatures. 
It is standard
practice to determine the  gravity  by imposing ionization balance on the 
\ion{Fe}{i} and \ion{Fe}{ii} lines. 
Unfortunately, with the relatively low S/N and  
moderate resolution of our spectra, very few \ion{Fe}{ii} lines were detectable, and they are all either weak features and/or in low S/N regions of the spectrum. We therefore  did not take into account  these \ion{Fe}{ii} lines and 
determined the gravity from the photometry, as explained above.
 
 For all the lines  belonging to the elements other than Fe, we used the  spectrum synthesis  
to determine the abundances.

\section { Error budget}

 Table \ref{errors} lists  an estimate of the errors due to typical uncertainties in the stellar parameters. These errors were estimated by varying T$_{eff}$, log g and $\xi$
  in the model atmosphere of SDSS~J163114.06+124526.6 by the amounts indicated in the table.  As the stars have similar stellar parameters, the other stars yield similar results. 
 The total error is estimated by adding  the quadratic sum of the the uncertainties in the stellar parameters and the error in the fitting procedure of the synthetic spectrum and the observed spectrum (the main source of error comes the incertitude in the placement of the continuum level). 

   \begin{table}
      \caption[]{Error Budget}
  \begin{tabular}{l c c c }
            \hline
            \noalign{\smallskip}
 Element             &    $\Delta T_\mathrm{eff}$    & $\Delta$ log g & $\Delta$ $\xi$  \\
                         &   $100~K $   &   $   0.3 dex$    & 0.3 km~s$^{-1}$  \\
                         \hline \\
Mg  &   0.12  & -0.05  & -0.12   \\
Al   &  0.12    & -0.10 &  -0.10   \\
ScII    & 0.07 & +0.05 & -0.11   \\
Ti    & 0.13     & -0.04  & -0.11   \\
Mn    & 0.15   & -0.06  &  -0.14  \\
FeI    & 0.15   & -0.06 & -0.12   \\
Ni    & 0.15     & -0.04 &  -0.15   \\
BaII  & 0.10    & +0.09  &  -0.11  \\
SrII    & 0.12   & +0.10  & -0.12  \\
          \noalign{\smallskip}
            \hline
         \end{tabular}
         \label{errors}\\
   \end{table}

\section {Results and discussion}

\begin{figure*}
   \centering
 \resizebox{\hsize}{!}{\includegraphics[clip=true]{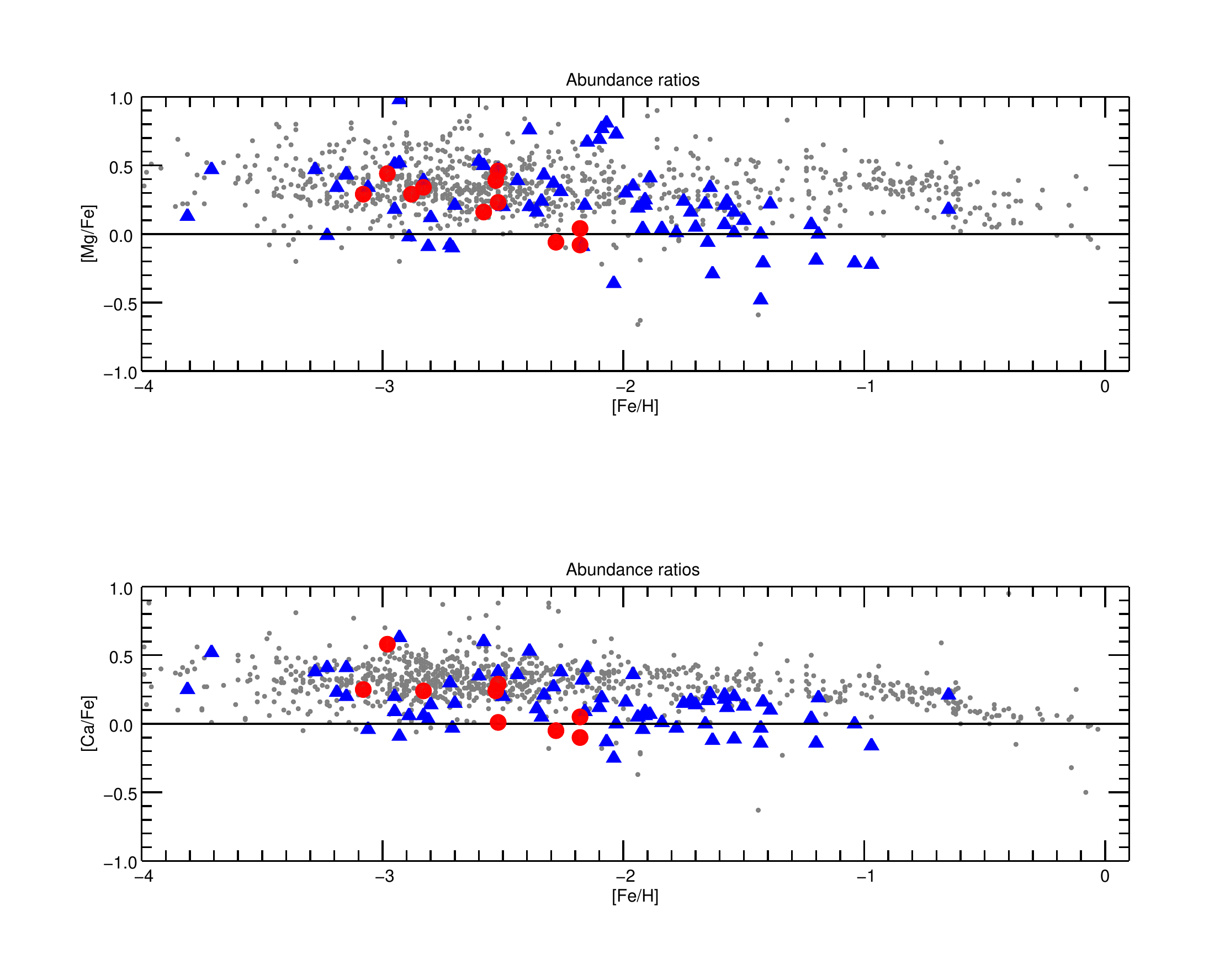}}
   \caption{Alpha elements : Grey circles represent literature data for field stars  gathered in  \citet{frebel2010} . Blue triangles are literature data for dwarf spheroidal galaxies. 
   Red circles represent the results for our sample of UfDSph stars.  }
              \label{Alpha}%
    \end{figure*}

The resulting abundances can be found in Table \ref{star_abund}.  Fig \ref{Alpha} presents the [Mg/Fe]  and [Ca/Fe] ratios found for our sample together  with literature data for  Milky Way field 
stars and stars in Dwarf Spheroidal galaxies.  The majority of the stars shows a high [Mg/Fe] ratio comparable to what is found in the halo stars. The [Ca/Fe]  ranges from $\simeq$ -0.05 dex for a star in CVn~I  stars to   $\simeq$ +0.65 dex for BooII stars.  This range is in agreement with the spread found for DSph stars, in the metallicity range $-$2.00 to $-$3.00 dex.  It is interesting to note that the [Mg/Fe] and [Ca/Fe] ratios  reach a solar value at a  metallicity lower than the Milky Way field stars (where it is reached at 
[Fe/H] $\simeq$ 0.0 dex) and than the dwarf galaxy stars for which the solar ratio is reached at metallicity around $-$2.00 to $-$1.50 dex in agreement with models of galactic chemical evolution \citep{vincenzo2014}.

\begin{figure*}
   \centering
 \resizebox{\hsize}{!}{\includegraphics[clip=true]{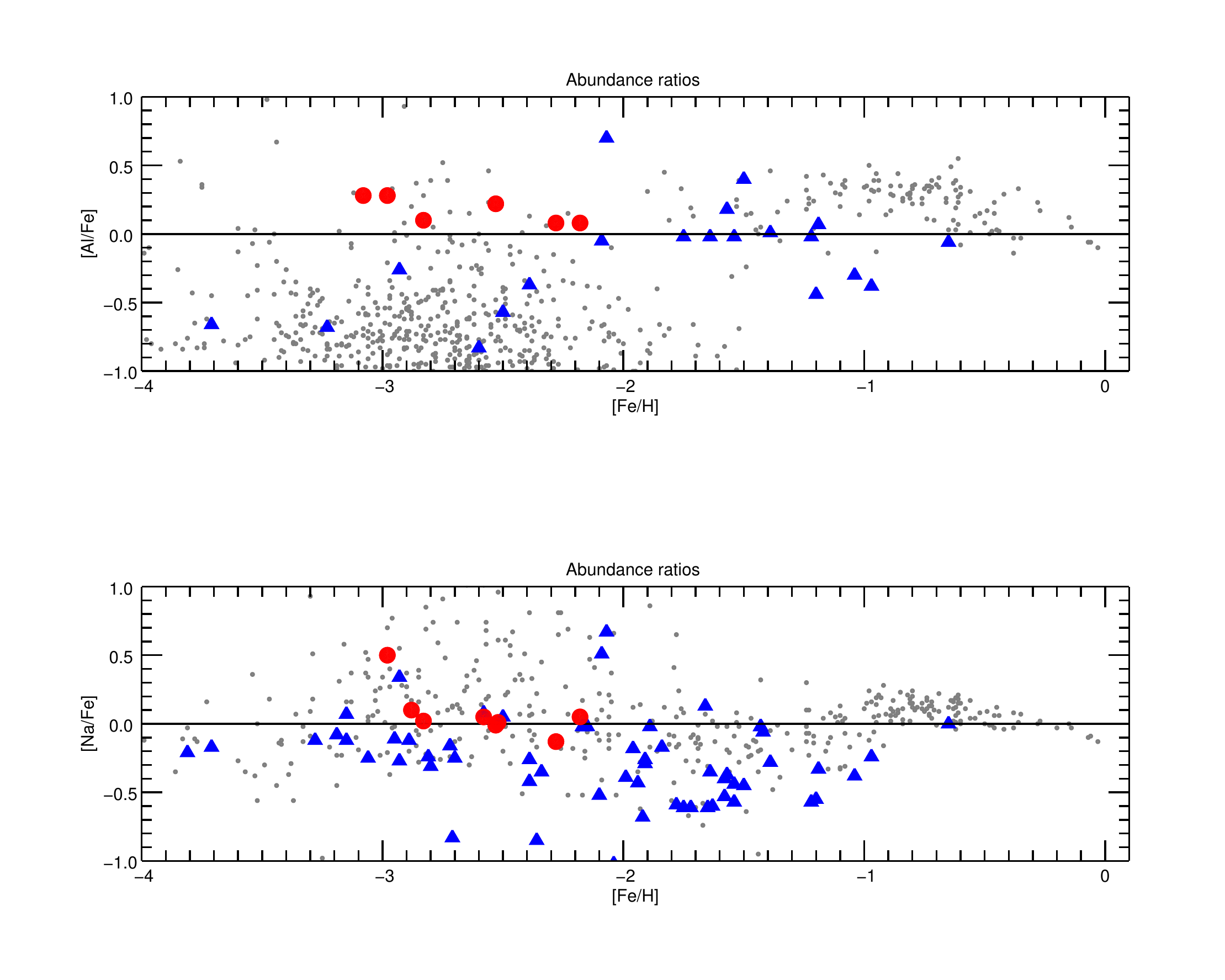}}
   \caption{ Al  and Na abundance ratios : Grey circles represent literature data for field stars  gathered in  \citet{frebel2010} . Blue triangles are literature data for dwarf spheroidal galaxies. 
   Red circles represent the results for our sample of UfDSph stars. }
              \label{AlFe}%
    \end{figure*}

The upper part of Fig \ref{AlFe} shows our results for the [Al/Fe] ratios. As for the previous figure, we have added literature data for field stars and dwarf spheroidal galaxy stars.  We find a high value of the 
[Al/Fe] ratio when compared to the halo stars.  We did not apply non-LTE corrections in order to make a direct comparison with the halo  and DSph stars which have been analyzed under the same assumptions. 
Sodium seems to share this behavior as shown in the lower part of Fig \ref{AlFe}. This high value of [Na/Fe] compared to the ratios in halo stars of the same metallicity has also been found by \citet{koch2008} in Hercules. 

\begin{figure*}
   \centering
 \resizebox{\hsize}{!}{\includegraphics[clip=true]{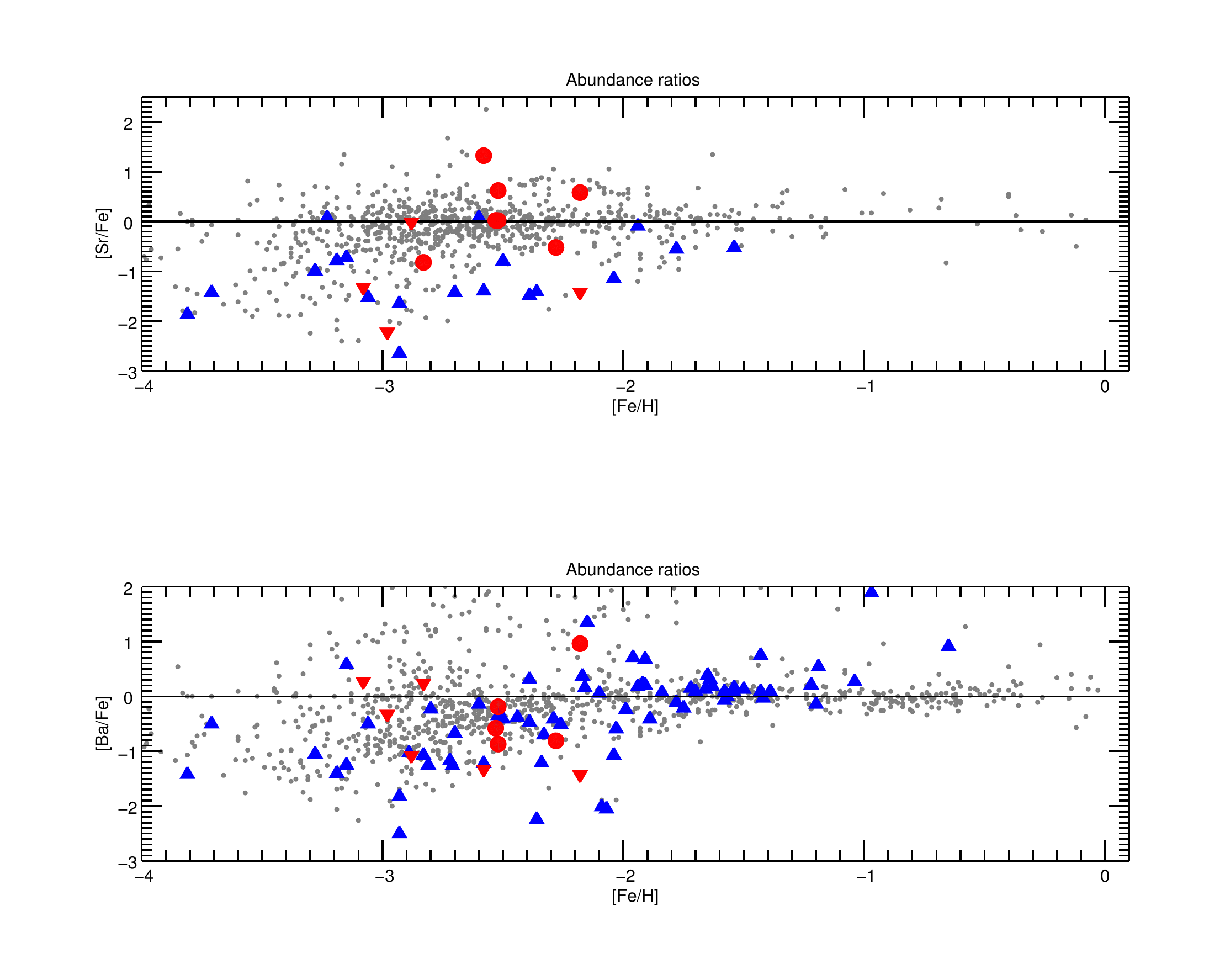}}
   \caption{Neutron capture  elements :  Grey circles represent literature data for field stars  gathered in  \citet{frebel2010}. Blue triangles are literature data for dwarf spheroidal galaxies.  Red circles represent the results for our sample of UfDSph stars.  Red triangles represent  upper limits for stars of our sample. }
              \label{ncapture}%
    \end{figure*}

Fig \ref{ncapture} presents the [Sr/Fe]  and [Ba/Fe] ratios found for our sample together  with literature data for field stars and dwarf spheroidal galaxy stars. The upper graph shows  a high value of the [Sr/Fe] ratio for the metal-rich sample of our stars similar to what is found in the halo stars and significantly different from the DSph stars. 
For the most metal-poor stars of our analysis, the ratio appears to be lower that  what found for the bulk of the field halo stars of the same metallicity. 
For Barium, our results fall also in the range of [Ba/Fe] found for halo stars.

   \begin{table*}
      \caption[]{Detailed abundances :  [X/Fe] for all the elements except Fe for which [Fe/H] is given.}
  \begin{tabular}{l r r r r r r   }
            \hline
            \noalign{\smallskip}

 Star                                                                &   C      &     Na~   &  Mg~ &    Al~  &   Ca~    &   Mn~           \\  
SDSS J163044.49+124947.8 / Her                &   ---     &   0.01  &    0.23  &   ---      &  -0.01   &  ---     \\  
SDSS J163059.32+124725.6 / Her                &   ---     &   0.02    &   0.34 &  0.15  &  0.29  &  ---      \\     
SDSS J163104.50+124614.4 / Her                &  -0.34 &   0.00   &   0.29 &  0.23   &  0.25  &  ---        \\     
SDSS J163114.06+124526.6 / Her                &   ---     &   -0.15   &   -0.06 &  0.18  &  0.05  &  ---       \\     
SDSS J113255.99-003027.8 / Leo IV -- S1   &   ---     &   0.10    &   0.44  &   ---    &   ---    &  ---       \\     
SDSS J113258.70-003449.9 / Leo IV            &  ---      & 0.05      &   -0.06  &  0.08 &  0.05  &  ---       \\     
SDSS J125713.63+341846.9 / CVn II           &  ---      &  -0.05    &   0.16  &   ---    &   ---    &  ---        \\     
SDSS J135801.42+125105.0 / Boo II -- 7     & -0.24   &    ---      &  0.29   &  0.28 &  0.25   &  0.04  \\     
SDSS J135751.18+125136.9 / Boo II -- 15   &  -0.10  &   0.50   &  0.44   &  0.28 &  0.58   & 0.44  \\     
SDSS J132844.25+333411.8 / CVn I            &   ---      &    ---     &  0.46   &   ---    &  0.29   &  ---    \\     
SDSS J132755.65+333324.5 / CVn I            &    ---     &    ---     &  0.04  &   ---    &  -0.10  &  ---      \\           
 \noalign{\smallskip}
 \hline
\noalign{\smallskip}

                                                                     &   Ti~      &   Fe~     &   Co~   &   Ni~      &    Sr~             &    Ba~       \\  
SDSS J163044.49+124947.8 / Her               &  ---         & -2.52  &  ---    &  ---     &        0.02    &       -0.87 \\  
SDSS J163059.32+124725.6 / Her               &  ---         & -2.83   &  ---   &  ---     &        -0.82    &  $<$  0.24 \\     
SDSS J163104.50+124614.4 / Her                &  ---        & -2.53  & 0.13  &  ---     &        0.08     &       -0.47 \\     
SDSS J163114.06+124526.6 / Her                &  ---        & -2.28  &  ---    & -0.24  &        -0.52    &       -0.81 \\     
SDSS J113255.99-003027.8 / Leo IV -- S1   & 0.26      & -2.88   &  ---    &  ---     &  $<$  -0.02 & $<$ -0.98 \\     
SDSS J113258.70-003449.9 / Leo IV              & -0.14  & -2.18   &  ---    &  ---    &  $<$ -1.42   & $<$ -1.38 \\     
SDSS J125713.63+341846.9 / CVn II               &  ---     & -2.58  &  ---    &  ---     &        1.32     & $<$ -1.28 \\     
SDSS J135801.42+125105.0 / Boo II -- 7      &  ---       & -3.08   &  ---    &  ---     &  $<$ -1.32   & $<$ 0.32 \\     
SDSS J135751.18+125136.9 / Boo II -- 15    &  ---       & -2.98   &  ----   &  ----   &  $<$ -2.22   & $<$ -0.28 \\     
SDSS J132844.25+333411.8 / CVn I             &  ---       & -2.52   &  ---    &  ---     &        0.62     &       -0.14 \\     
SDSS J132755.65+333324.5 / CVn I               &  ---    & -2.18   &  ---    &  ---     &        0.58     &       0.36 \\

            \noalign{\smallskip}
            \hline
         \end{tabular}
         \label{star_abund}\\
   \end{table*}

\subsection {Bootes II}

We have observed two stars in the Galaxy (BooII-7 and BooII-15) and found metallicities [Fe/H] = $-$2.98 dex and $-$3.08 dex.  As the second star has been already observed by 
\citet{koch2014}, we put the results from both studies in table \ref{star15_abund}. The results are in general good agreement.   
The Carbon abundance has been computed by fitting a synthetic spectrum for the CH G band  
The [$\alpha$/ Fe]  overabundance  and the low [Ba/Fe] are characteristic of the galactic halo population. 
We found a very low upper limit for strontium with a  value of  [Sr/Fe] $\le$ $-$2.22 dex.  We also obtained a comparable 
low value of strontium for the star Boo-7 with  [Sr/Fe] $\le$ $-$1.32 dex.   This low value of strontium  with respect to what is found in the halo stars of the same metallicity
is generally observed in UfDSph galaxies as shown in Fig \ref{ncapture}.

   \begin{table}
      \caption[]{BooII-15 abundance comparison}
  \begin{tabular}{l c c}
   \hline
   \noalign{\smallskip}
Ion                 &   This paper          &     Koch \& Rich (2014)  \\  
$[$Fe/H$] $               &     -3.08  &   -2.93  \\
$[$C/Fe$] $         &    -0.10  &    0.03     \\  
$[$Mg/Fe$] $          &       0.44  &   0.58      \\  
$[$Ca/Fe$] $          &       0.58  &    0.35     \\  
$[$Ba/Fe$] $  & $<$ -0.28  &  $<$  -0.62     \\  
            \noalign{\smallskip}
            \hline
         \end{tabular}
         \label{star15_abund}\\
   \end{table}

 \subsection {Canes Venatici I}
\begin{figure}
   \centering
 \resizebox{\hsize}{!}{\includegraphics[clip=true]{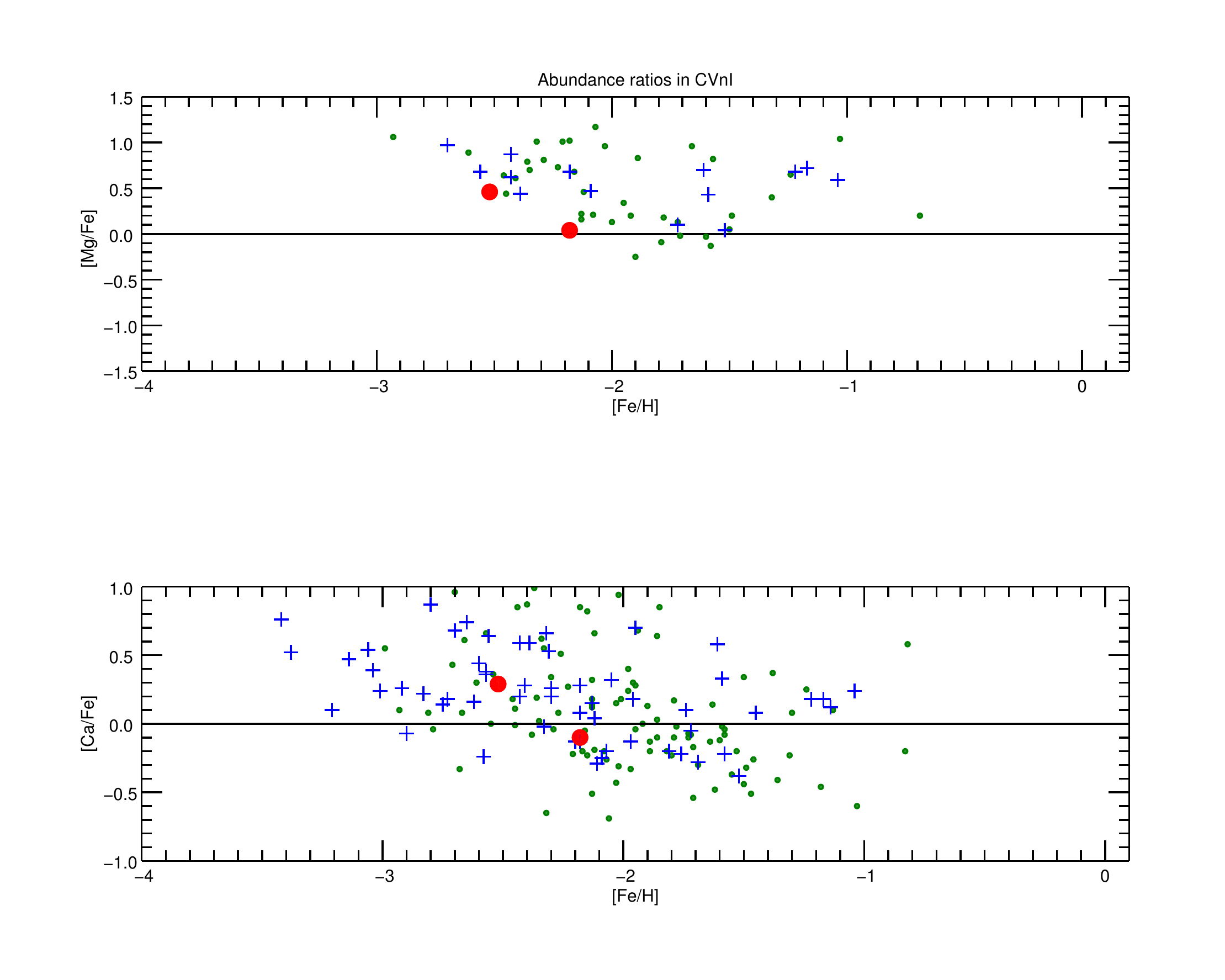}}
   \caption{Abundance results for  the CVnI  galaxy  stars. Red circle represent our two stars. Green symbols are results from \citet{kirby2010}. Blue symbols are results from \citet{vargas2013} }
              \label{CVnI_alpha}%
    \end{figure}

 Abundances of Fe, Mg and Ca of a sample of stars belonging to  CVnI have been reported by \citet{kirby2010} using low resolution spectra. Using  the same Keck/DEIMOS medium resolution spectra obtained by \citet{kirby2010},  \citet{vargas2013} determine the abundance of Fe, Mg, Ca in stars of this galaxy.
On Fig \ref{CVnI_alpha}, we plotted  our results together with the results from  \citet{kirby2010} and \citet{vargas2013}. We have also added the results for a sample of UfDSph galaxies as in 
Fig \ref{Alpha}.

We report for the first time the abundance determination of  the neutron capture elements in two stars of this galaxy.   We found for both stars a  high ratio of [Sr/Ba] , +0.22 dex and +0.76 dex respectively.  It is interesting to note that the these ratios are similar to the one found in the halo stars at the same [Ba/H] abundance as found by \cite{francois2007}. 

 \subsection {Canes Venatici II}

\begin{figure}
   \centering
 \resizebox{\hsize}{!}{\includegraphics[clip=true]{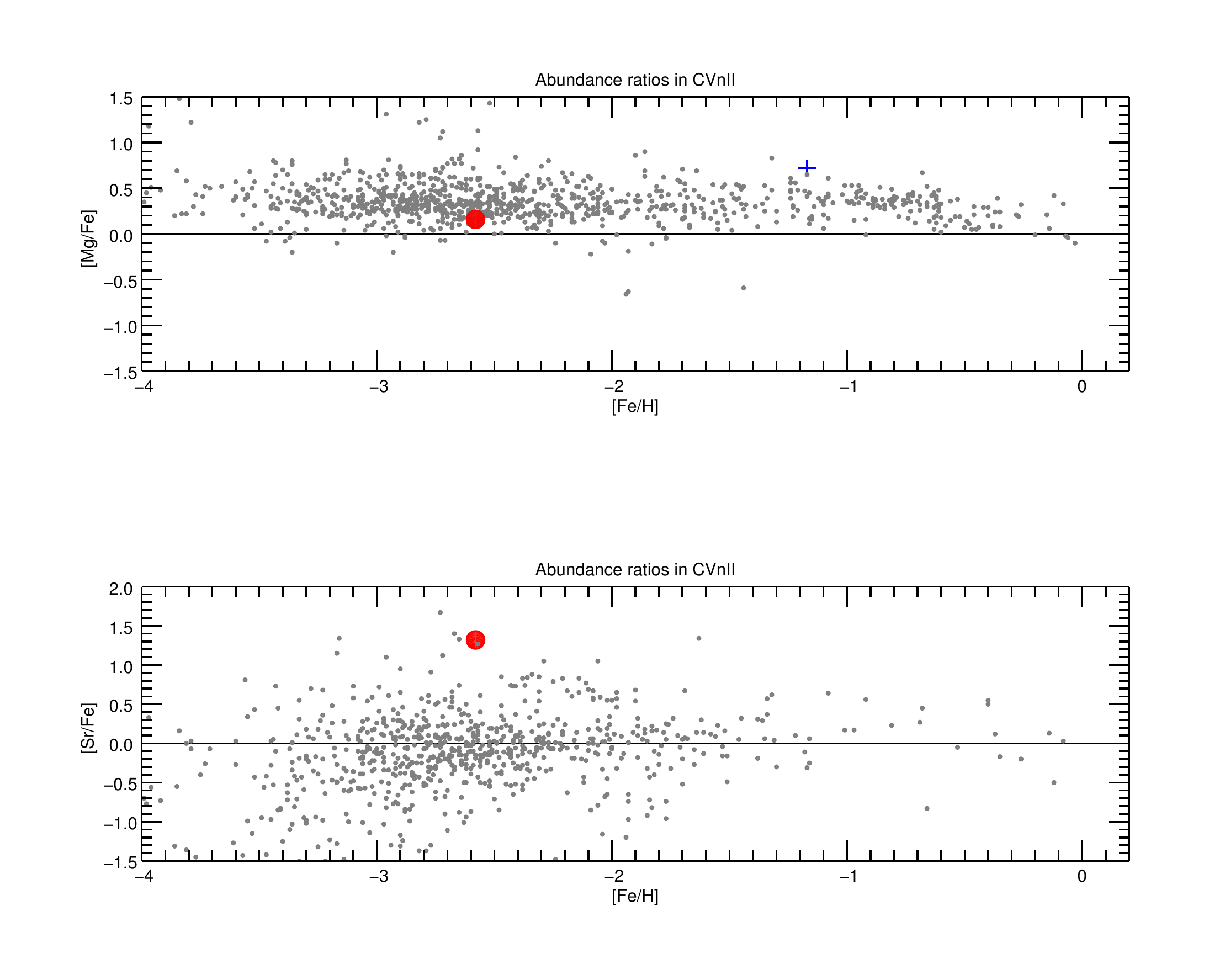}}
   \caption{Abundance results for  the CVnII  galaxy  stars. Red circles represent the abundance results for  our  stars.    }
              \label{CVnII}%
    \end{figure}

Our results for   Canes Venatici II  are presented in Fig \ref{CVnII}.   This is the first Strontium abundance determination for a star in this Galaxy. 
Our star has a very high [Sr/Fe] values and  and a low upper limit of [Ba/Fe] which makes it a star with a exceptionally high [Sr/Ba]  with a value larger than 2.6 dex.  

\begin{figure}
   \centering
 \resizebox{\hsize}{!}{\includegraphics[clip=true]{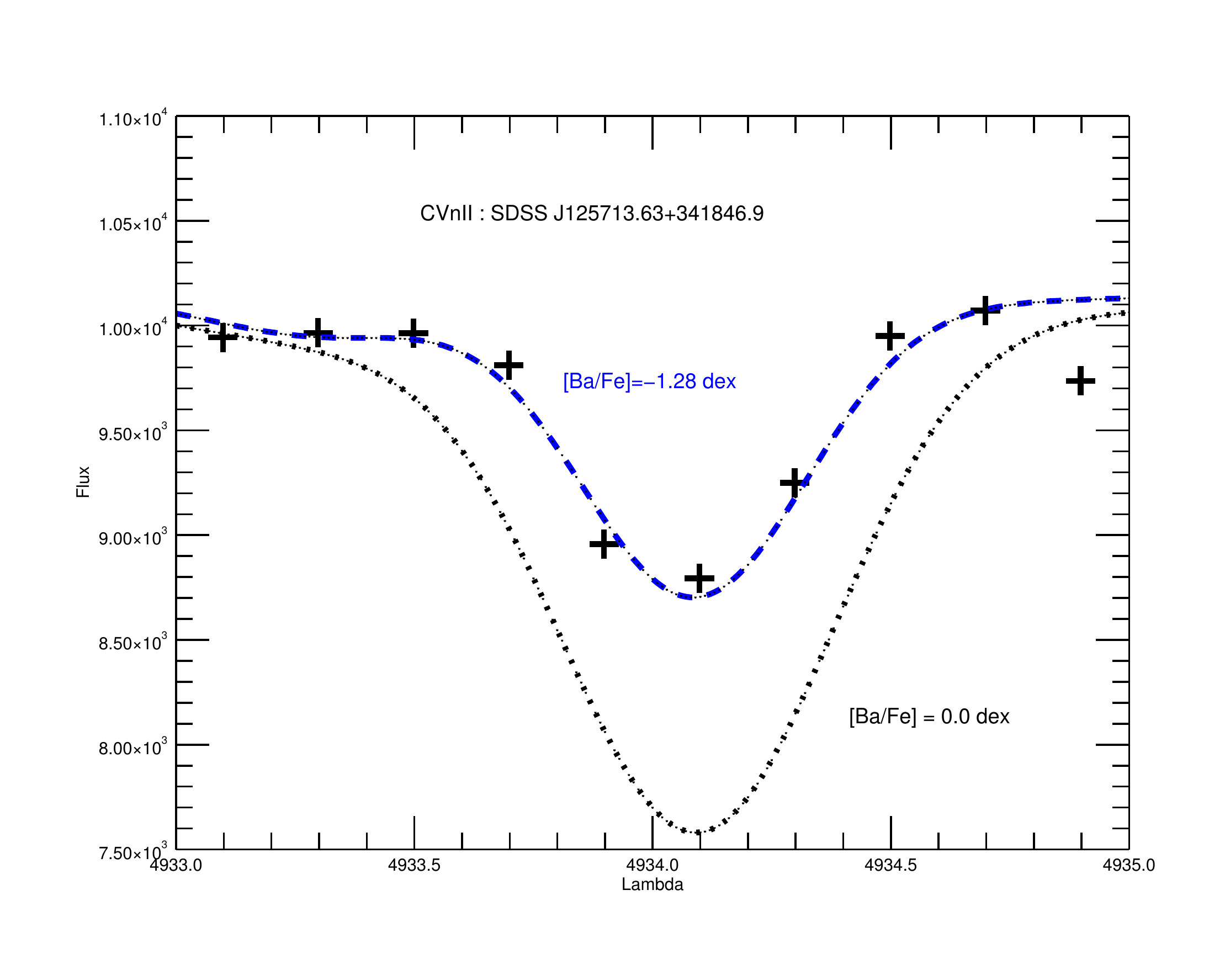}}
   \caption{ Comparison of the observed spectrum represented by pluses and synthetic spectra with different barium abundances. }
              \label{Ba4934}%
    \end{figure}

\begin{figure}
   \centering
 \resizebox{\hsize}{!}{\includegraphics[clip=true]{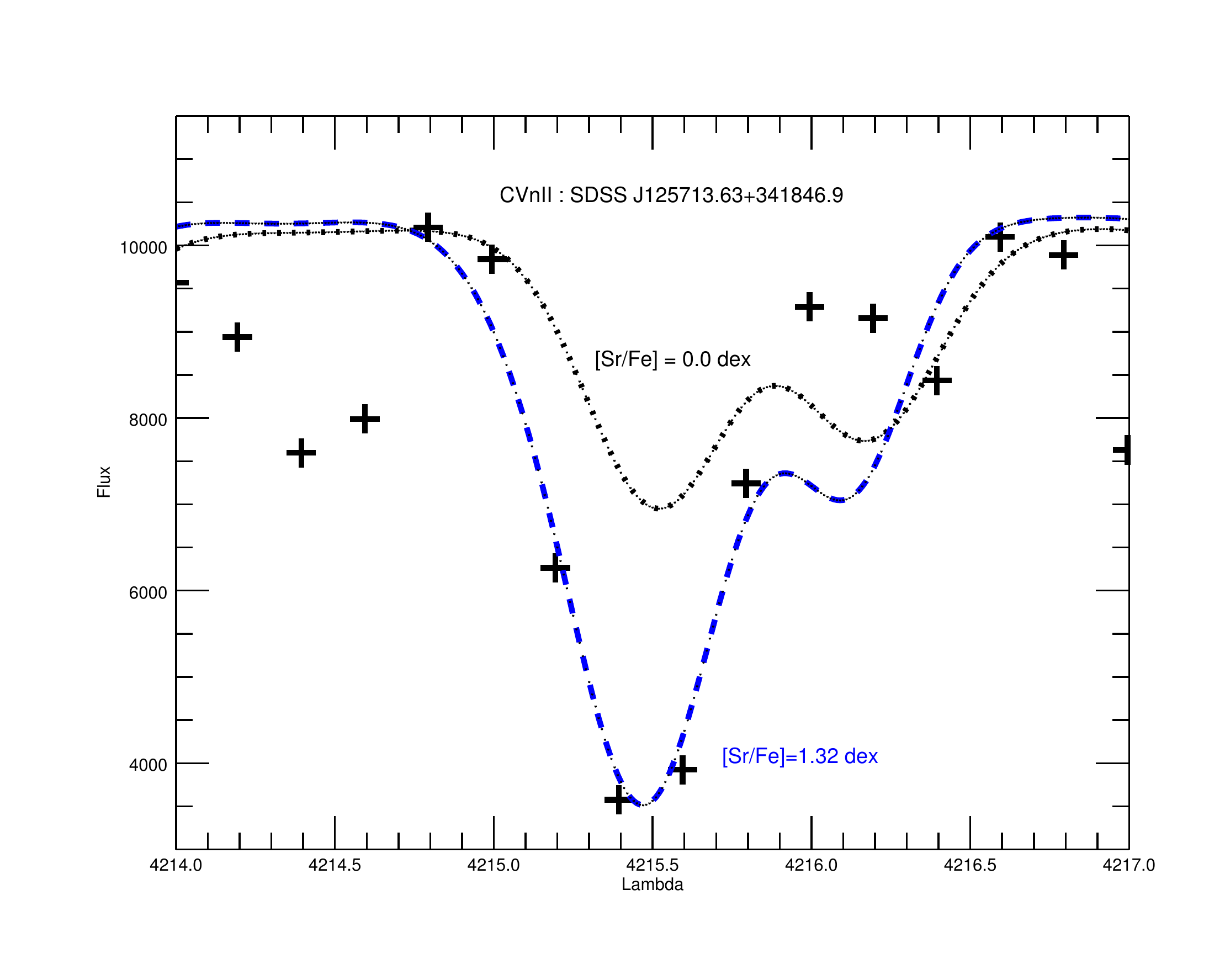}}
   \caption{Comparison of the observed spectrum represented by pluses and synthetic spectra with different strontium abundances. }
              \label{Sr4215}%
    \end{figure}

 In Figs \ref{Ba4934} and \ref{Sr4215} are shown the results from the spectrum synthesis computation superimposed on the data  for the line of Barium at 493.4 nm and the line of strontium at 421.5 nm. The blue lines correspond to the abundance ratios we determined whereas the black dotted line represents  a spectrum with a solar ratio. 
 The low barium abundance has been confirmed using the lines at 649.7 and 614.1 nm. 
The high [Sr/Ba] ratio may be explained by invoking different sources
 for the production of light neutron capture elements versus heavier neutron capture elements.  
  At low metallicity, strontium may be formed by the weak r-process  \citep{wanajo2013}.
 The large difference between the two mostly s-process element strontium and barium may come from a peculiar pollution of the cloud which formed the star, the source  being possibly a core-collapse supernova  as proposed by \citet{wanajo2013}.  More recently, \citet{cescutti2015} have computed detailed models of galactic chemical 
 evolution of our Galaxy.  Their computations have shown that the combination of r-process production by neutron star mergers and s-process by spinstars  \citep{pignatari2008,frischknecht2012} is able to reproduce the large range of [Sr/Ba]  ratios at low metallicity.
  
It would be particularly interesting to obtain a high resolution high S/N spectrum of this star in order to detect and measure the abundances of other n-capture elements and compare it with high Sr low metallicity  field halo stars.  

 \subsection {Hercules}

\begin{figure*}
   \centering
 \resizebox{\hsize}{!}{\includegraphics[clip=true]{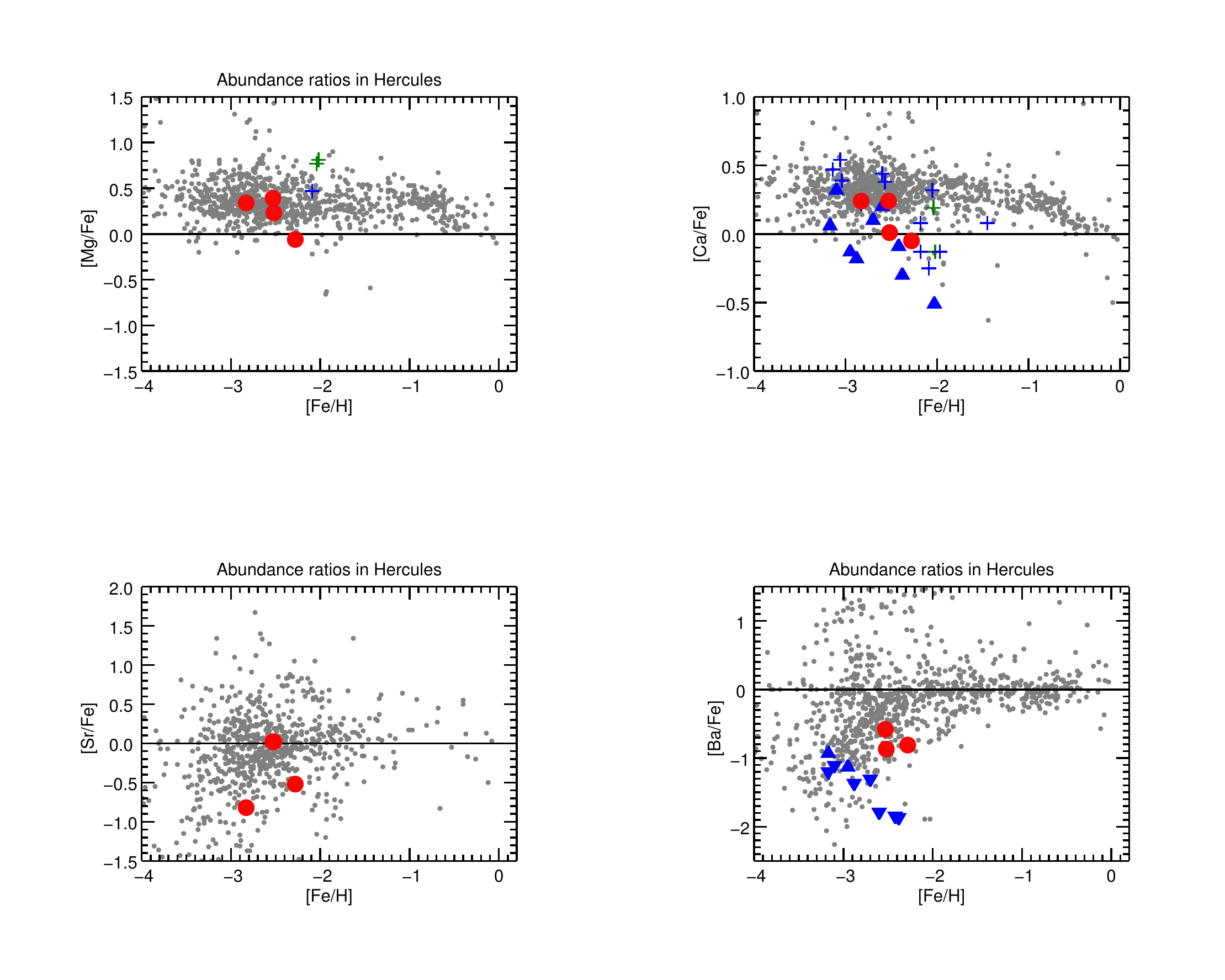}}
   \caption{Our results for Hercules are presented as red circles. We have added the results from \citet{koch2008,koch2013} as blue triangles (triangles pointing down are
   upper limits)  and \citet{aden2009} as blue pluses. Grey dots are literature data for the field halo stars  gathered in  \citet{frebel2010}. }
              \label{Her_alpha}%
    \end{figure*}

\citet{koch2013} studied  a sample of 11 red giant stars. They could detect the barium line at 6141.713 $\AA$ for three of them. 
Our results for Hercules are presented as red circles in Fig \ref{Her_alpha}. We have added the results from \citet{koch2008,koch2013}  and \citet{aden2009}.

 \begin{figure}
   \centering
 \resizebox{\hsize}{!}{\includegraphics[clip=true]{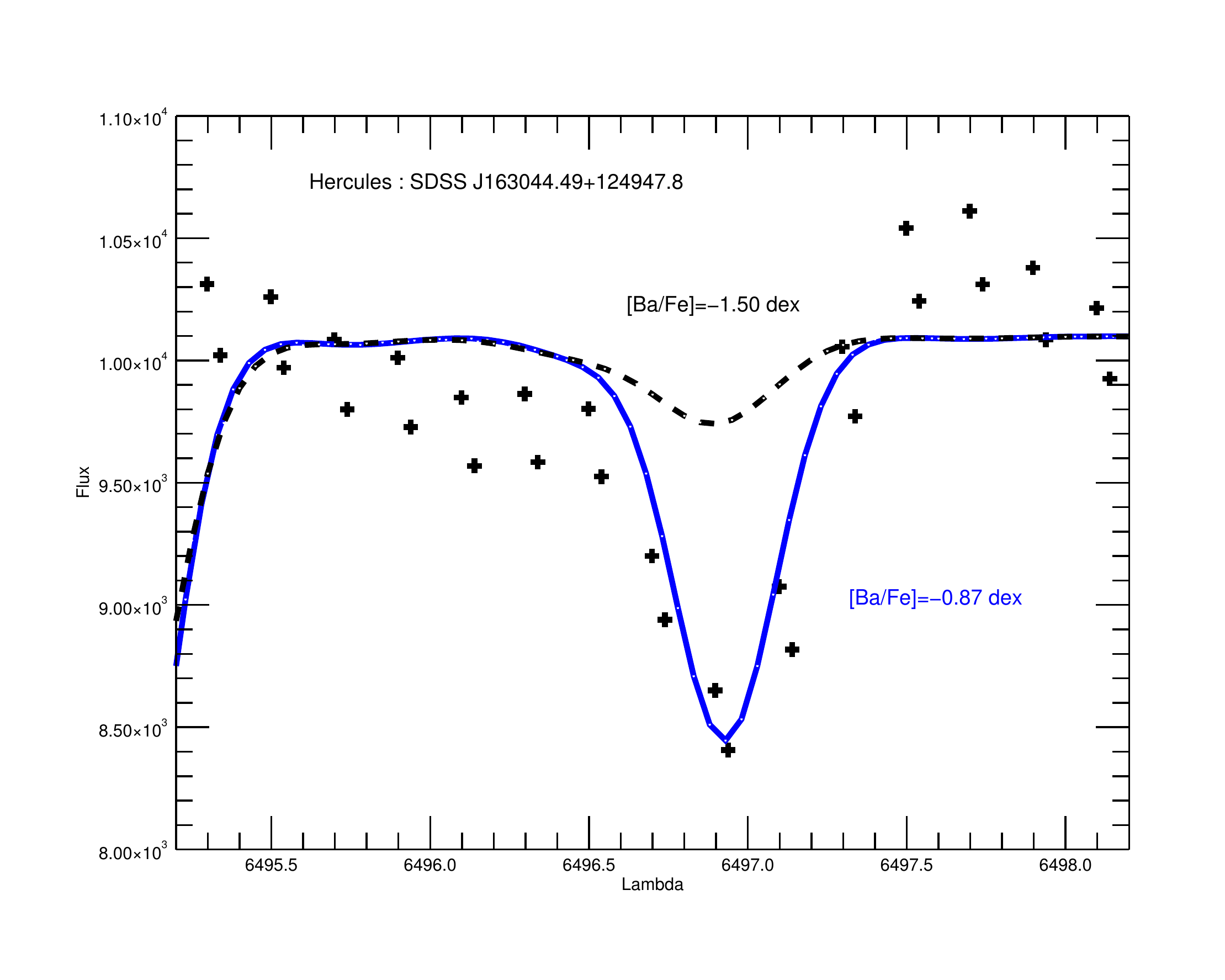}}
   \caption{Comparison of the observed spectrum represented by pluses and synthetic spectra. The dashed black lines is computed for a barium abundance   [Ba/Fe] = -1.50 dex corresponding to the 3 $\sigma$ upper limit of \citet{koch2013}. The higher quality of the X-Shooter spectra permits the detection of the barium line.  The blue line is computed with a barium value  [Ba/Fe] = -0.87 dex  corresponding to our result.  }
              \label{BaHer400508}%
    \end{figure}

  \begin{figure}
   \centering
 \resizebox{\hsize}{!}{\includegraphics[clip=true]{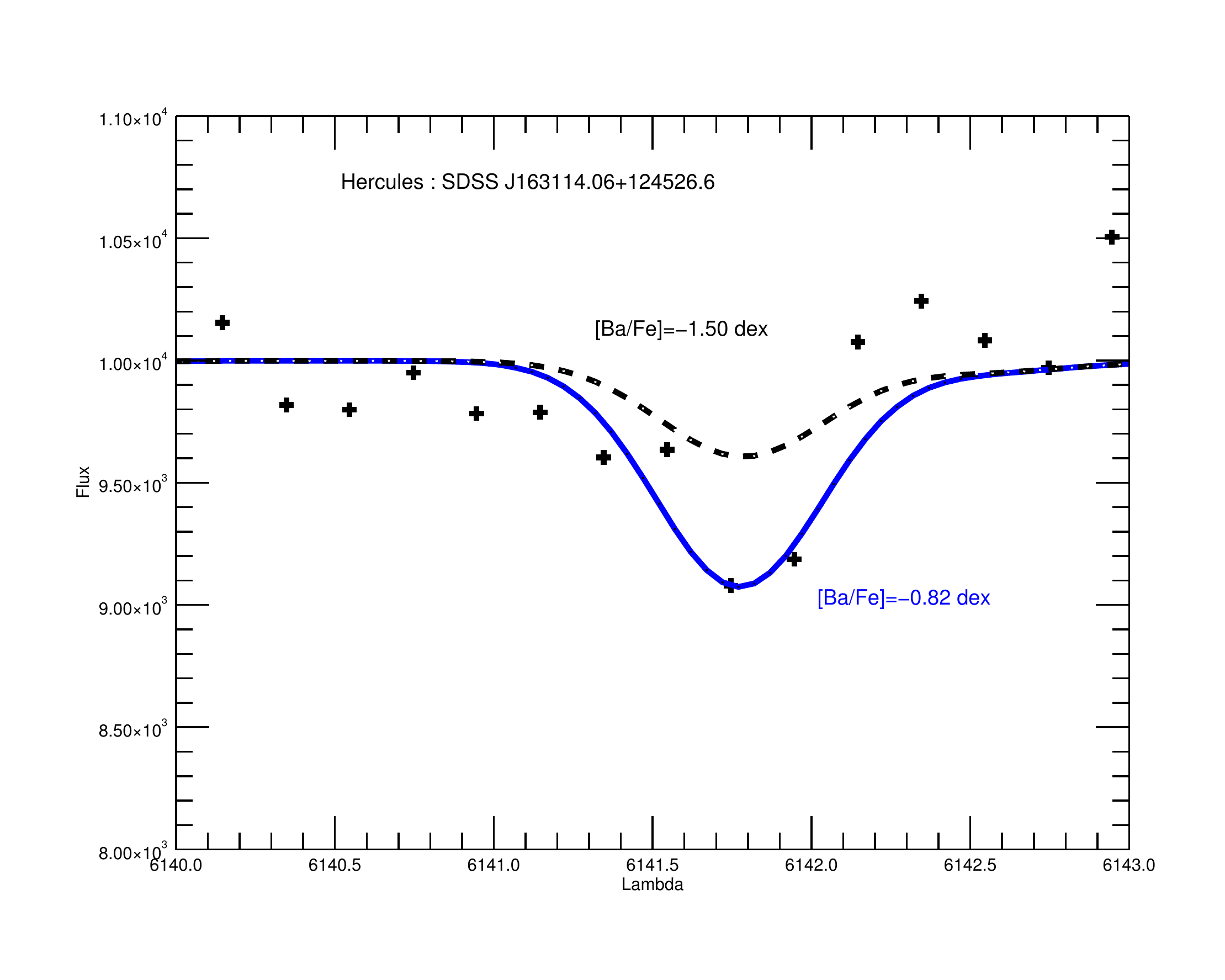}}
   \caption{Comparison of the observed spectrum represented by pluses and synthetic spectra. The dashed black lines is computed for a barium abundance   [Ba/Fe] = -1.50 dex corresponding to the 3 $\sigma$ upper limit of \citet{koch2013}. The higher quality of the X-Shooter spectra permits the detection of the barium line.  The blue line is computed with a barium value  [Ba/Fe] = -0.81 dex  corresponding to our result.    }
              \label{BaHer466011}%
    \end{figure}

Our sample has  metallicities ranging from  $-$2.28 dex to $-$2.83 dex.  
Our results show clearly an increase of the  [$\alpha$/Fe]
  ratios as the metallicity decreases.   It is important to note  that this effect has been already observed by \citet{vargas2013} not only  Hercules but also in other  galaxies.
This is what is expected with classical models of chemical evolution where the impact of the contribution of type SNIa iron  
on the abundance ratios [$\alpha$/Fe] vs metallicity
relations is shown as a decrease of this ratio as the metallicity increases.  For Her, the solar ratio is reached at a much  lower metallicity that the one found for the Milky Way and even the 
dwarf spheroidal galaxies such as Carina or Sculptor as shown in \citet{vincenzo2014}. Our results for Calcium  are in good agreement with the results   \citet{aden2009}
although we notice a slightly higher $[$Ca/Fe$]$ ratio than the one found by  \citet{koch2008,koch2013} . 

 In Fig \ref{BaHer400508} and \ref{BaHer466011}, we show the spectrum synthesis of a  barium line with two assumptions for the [Ba/Fe] ratio. 
The high efficiency of X-Shooter allowed to make a clear  detection of the barium line  compared to previous studies where only upper limits could be derived.
 
For barium, the combination of our results with the barium detections from   \citet{koch2008,koch2013} seem to indicate an increase of the [Ba/Fe] ratio as the metallicity increases in line 
with what is found in our Galaxy.

However, this should be taken with caution when we add their  Ba upper limits as it would rather reveal a large scatter.

 \subsection {Leo IV}

     \begin{table}
      \caption[]{Leo IV - S1 abundance ratio comparison}
  \begin{tabular}{l c c}
   \hline
   \noalign{\smallskip}
Ion                  &   This paper          &     Simon et al. (2010)  \\  
$[$Fe/H$] $     &   -2.88     &     -3.20            \\
$[$Na/Fe$] $     &    0.10   &   0.01                \\  
$[$Mg/Fe$] $   &     0.44   &           0.32       \\  
$[$Ti/Fe$] $    &  0.26      &             0.38    \\
$[$Sr/Fe$] $    &  $<$ -0.02     &    -1.02       \\
$[$Ba/Fe$] $    &   $<$ -0.98     &              -1.45   \\  
            \noalign{\smallskip}
            \hline
         \end{tabular}
         \label{star_S1_abund}\\
   \end{table}

We observed two stars in Leo IV , one of them has been already studied by \citet{simon2010}. In Table \ref{star_S1_abund}, we can compare the results from both studies. 
The results are in good agreement.  The high resolution spectrum used by \citet{simon2010} allowed to derive the abundance of Ba and Sr. For our second star, we found a 
higher metallicity with $[Fe/H]$ = $-$2.18 dex,  $[Mg/Fe]$ = -0.06 dex and $[Ca/Fe]$ = $-$0.05 dex in good agreement with the theoretical predictions from the galactic chemical evolution models of \citet{vincenzo2014}

\subsection{ Do  fossil galaxies have peculiar abundances  ?}

    \begin{figure}
   \centering
    \resizebox{\hsize}{!}{\includegraphics[clip=true]{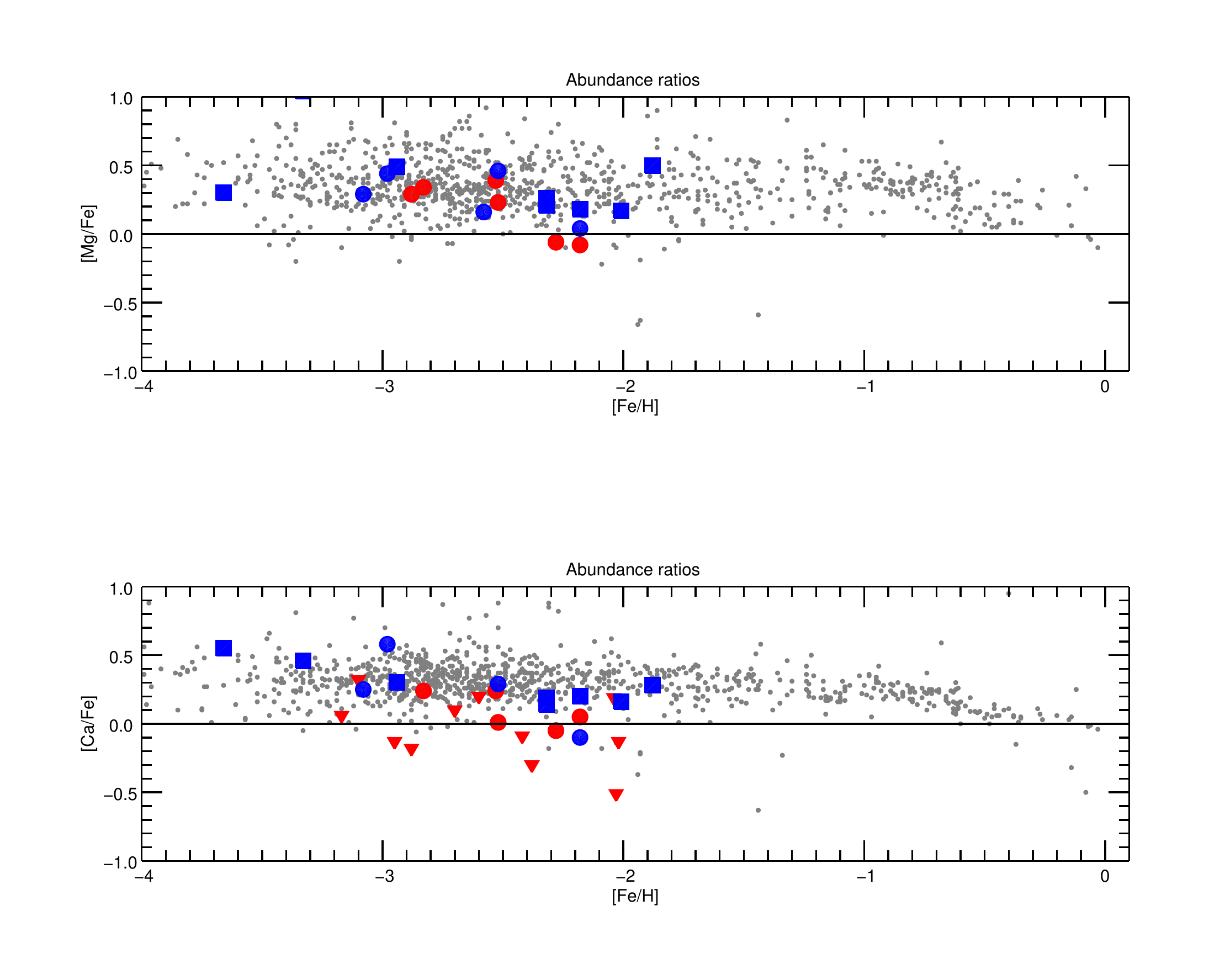}}

   \caption{Our results for Hercules and LeoIV  are presented as red circle . We have added the results from \citet{koch2008,koch2013} and \citet{aden2009} as red triangles (triangles pointing down are
   upper limits) . Blue circles are our results for the  remaining galaxies. Blue squares are literature data for BooI.  Grey dots are literature data for the field halo stars  gathered in  \citet{frebel2010}. }
              \label{Quenched_alpha}%
    \end{figure}

   \begin{figure}
   \centering
    \resizebox{\hsize}{!}{\includegraphics[clip=true]{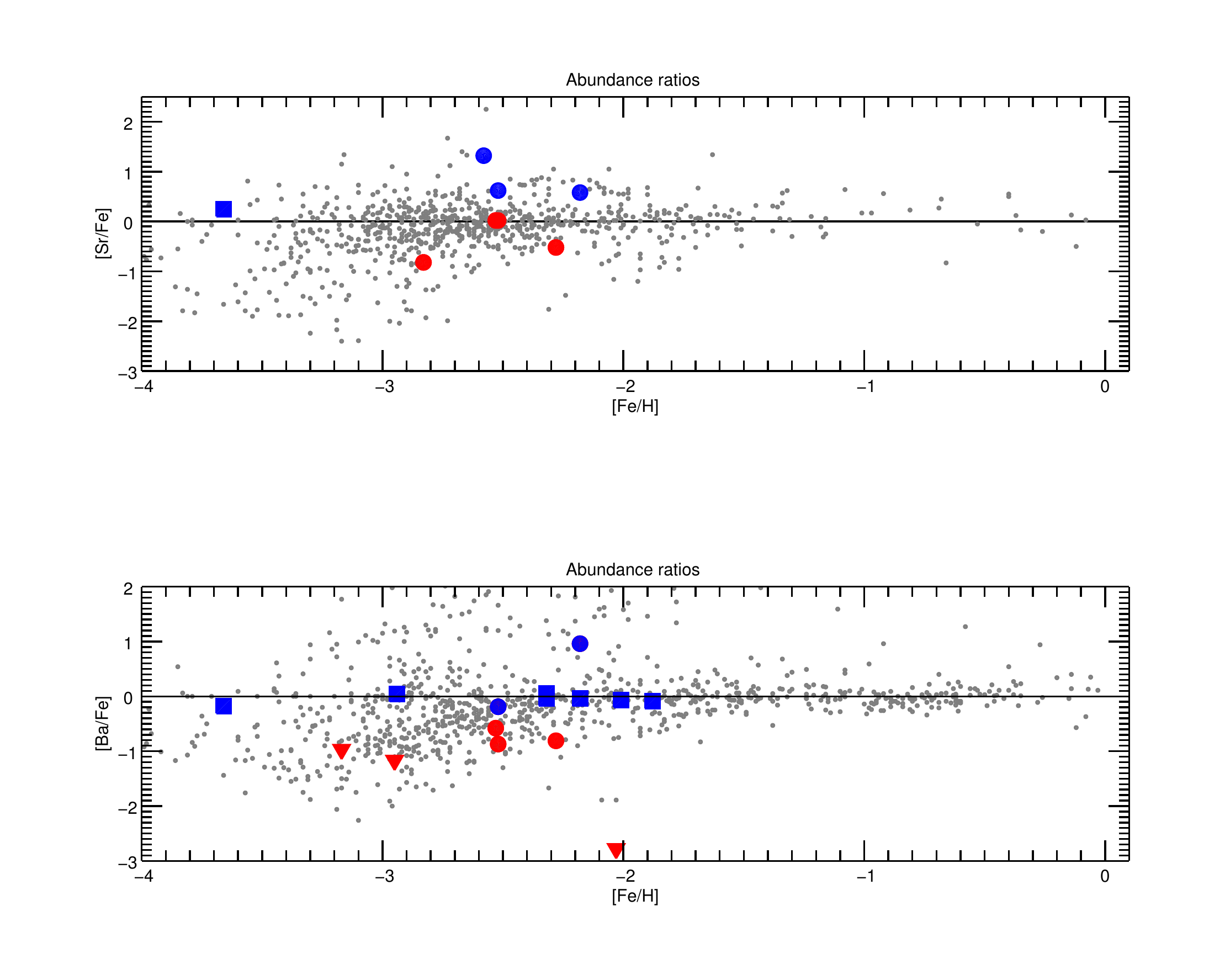}} 
   \caption{Our results for Hercules and  LeoIV  are presented as red circle . We have added the results from \citet{koch2008,koch2013} and \citet{aden2009} as red triangles (triangles pointing down are
   upper limits) . Blue circles are our results for the  remaining galaxies. Blue squares are literature data for BooI.  Grey dots are literature data for the field halo stars  gathered in  \citet{frebel2010}. }
              \label{Quenched_neutrons}%
    \end{figure}

 Among the fives galaxies studied in this paper,  two (Her and Leo IV  )  have probably formed   the bulk of their stars before reionization \citep{weisz2014}  
It would be therefore be particularly interesting to check whether the abundance ratios reveal any systematic difference between these "fossil" galaxies and the rest of the sample.   We have added the results from BooI \citep{gilmore2013} as a member of the galaxy group with an extended star formation history. 
 
In Fig \ref{Quenched_alpha} , we plotted our results and literature data for UfDSph  with red symbols for the fossil galaxies  and blue symbols for the other galaxies.  We also added as reference literature data for the field halo stars as small grey circles.   
An inspection of the figure seems to indicate the "fossil " galaxies  have a  lower $[$Ca/Fe$]$ than the other galaxies and
that $[$Mg/Fe$]$ is also somewhat lower.
In Fig \ref{Quenched_neutrons}, we made similar plots for the neutron capture elements Sr and Ba. Again, the "fossil" galaxies seem to have a
lower [Sr/Fe] and [Ba/Fe] than the other galaxies.  Only the high [Sr/Fe] found  in the CVnII star departs form this trend. 
 This result has to be taken with caution as it relies on a small number of stars.  Further studies based on a larger sample of galaxies would be necessary to confirm  the reality of this effect. 

On the assumption the "fossil" group of galaxies have indeed
produced the bulk of their stars before  reionization,  
our results, combined with the  literature data, 
suggest that  fossil galaxies have lower $[X/Fe]$  ratios  at any given metallicity, 
than the galaxies that have {\bf not} experienced a discontinuity  in their SFR (quenching).

The star formation history of quenched galaxies is affected by an episode when the formation of stars is stopped. In terms of galactic chemical evolution, this can be translated by a period where the 
intermediate mass stars continue their evolution while no  stars are formed. 
These stars are responsible for the enrichment in $s-process$ elements,
such as Sr and Ba.

 \section{Conclusions}

We have reported abundance ratios in a sample of  11 stars belonging to 5 different UfDSphs based on X-Shooter spectra obtained at the VLT. 
This study demonstrates that X-Shooter is a very powerful instrument  to determine the detailed  chemical composition of   metal-poor stars in UfDSph.  With the present analysis  based on only a couple of nights of telescope time, we could obtain some interesting results.
We can therefore foresee further studies  of  the detailed chemical evolution of many galaxies of the local group using 10 meter class telescopes and 
medium resolution spectroscopy at the level of R $ \simeq$ 8 000 and higher. 
 From the comparative analysis of the abundances ratios found in these different systems, we can not only study the star formation histories of these galaxies as entities but we can also check for the universality of the nucleosynthesis of the elements. In particular,  UfDsph (low mass galaxies) are ideal  to study the existence and the frequency  of rare events like neutron stars mergers and their impact on nucleosynthesis and galactic chemical evolution. 

Considering all the stars as representative of the same population of low mass galaxies, we found that the [$\alpha$/Fe] ratios  vs [Fe/H]  decreases as the metallicity of the star
increases  in a way similar to what is found for the population  of stars belonging to dwarf spheroidal galaxies.  The main difference is that the a solar [$\alpha$/Fe] is reached at a much lower 
metallicity for the UfDSph than for the dwarf spheroidal galaxies.   [Al/Fe] and [Na/Fe]  seem to be give higher values compared to the stars with the same metallicity observed in the halo or in dwarf spheroidal galaxies.

We report for the first time the abundance of strontium in CVnI.  The star we analyzed in this galaxy has a very high [Sr/Fe] and a very low 
upper limit of barium which makes it a star with an exceptionally high [Sr/Ba] ratio. 

Based on our results, we suggest that fossil galaxies, 
that have formed the bulk of their stars before 
reionization have lower 
$[X/Fe]$  ratios  than galaxies, of the same
metallicity, that have experienced a quenching of their
star formation rate.

%
%

 %

\begin{acknowledgements}
We would like to thank E.  Kirby for sending in electronic format the data for the individual stars he studied in his 2008 paper.
PF thanks  the European Southern Observatory for his support.   
PF and PB acknowledge support from the
Programme National de Physique Stellaire (PNPS) of
the Institut National de Sciences de l'Univers of CNRS. 
LM acknowledges support from 'Proyecto interno'  of the Universidad Andr\'es Bello.
CMB acknowledges support from FONDECYT regular project 1150060.
This research has made use of NASA's Astrophysics Data
System, and of the VizieR catalogue access tool, CDS, Strasbourg, France
      \end{acknowledgements}



\begin{thebibliography}{}

\bibitem[Ad\`en et al. (2009)]{aden2009}Ad\' en, D., Feltzing, S., Koch, A., Wilkinson, M. I., Grebel, E. K., Lundstrom, I., Gilmore, G. F., Zucker, D. B., Belokurov, V., Evans, N. W., Faria, D. 2009 \aap, 525, 153

\bibitem[Ad\`en et al. (2011)]{aden2011}Ad\' en, D., Eriksson, K., Feltzing, S., Grebel, E. K., Koch, A., Wilkinson, M. I. 2011, \aap, 525,153

\bibitem[Allen et al. (2012)]{allen2012} Allen, D., Ryan, S;G., Rossi, S., Beers, T., Tsangarides, S.A. 2012, \aap 548, 34

\bibitem[Alonso et al.(1999)]{alonso1999} Alonso, A., Arribas, S., \& Mart{\'{\i}}nez-Roger, C.\ 1999, \aaps, 140, 261 

\bibitem[Alvarez  \& Plez (1998)]{alvarez1998}  Alvarez, R., \& Plez, B. 1998, \aap, 330, 1109

\bibitem[Asplund et al. (1997)]{asplund1997} Asplund, M., Gustafsson, B., Kiselman, D., \& Eriksson, K. 1997, \aap, 318,
521

\bibitem[Asplund et al. (2009)]{asplund2009} Asplund, M.,Grevesse, N., Sauval, A. J.; Scott, P. \araa 47, 481

\bibitem[Belokurov et al.(2007)]{belo2007}  Belokurov, V., Zucker, D. B.,Evans, N. W.,Kleyna, J. T.,Koposov, S.,Hodgkin, S. T.,Irwin, M. J.,Gilmore, G.,Wilkinson, M. I.,Fellhauer, M.,Bramich, D. M.,Hewett, P. C.,Vidrih, S.,De Jong, J. T. a.,Smith, J. a.,Rix, H. -W.,Bell, E. F.,Wyse, R. F. G.,Newberg, H. J.,Mayeur, P. a.,Yanny, B.,Rockosi, C. M.,Gnedin, O. Y.,Schneider, D. P.,Beers, T. C.,Barentine, J. C.,Brewington, H.,Brinkmann, J.,Harvanek, M.,Kleinman, S. J.,Krzesinski, J.,Long, D.,Nitta, A.,Snedden, S. a. 2006, \apj,  654, 906 

\bibitem[Brown et al. (2014)]{brown2014} Brown, Thomas M., Tumlinson, Jason, Geha, Marla, Simon, Joshua D.,  2014  \apj,796, 91

\bibitem[Bullock et al. (2000)]{bullock2000}Bullock, J. S., Kravtsov, A. V., Weinberg, D. H. 2000, \apj, 539, 517

\bibitem[Caffau et al. (2011)]{caffau2011} Caffau, E., Bonifacio, P., Fran\c cois, P. et al. 2011 \aap, 542, 4

\bibitem[Caffau et al. (2011b)]{caffau2011b} Caffau, E.,  Ludwig, H.-G., Steffen, M., Freytag, B., Bonifacio, P.  2011 Solar Phys 268, 255 

\bibitem[Cescutti et al. (2015)]{cescutti2015} Cescutti, G., Romano, D., Matteucci, F., Chiappini, C., Hirshi, R. 2015, \aap 577, 139

\bibitem[Coleman et al.  (2007)]{cole2007} Coleman, M. G., de Jong, J. T. , Martin, N. F. , 2007 \apj, 668, 43

\bibitem[D'Odorico et al. (2006)]{dodo2006}D'Odorico, S., Dekker, H., Mazzoleni, R. et al. 2006 SPIE 6269, 33

\bibitem[Edvardsson et al. (1993)]{edvardsson1993} Edvardsson, B., Andersen, J., Gustafsson, B., et al. 1993, \aap, 275, 101

\bibitem[Fran\c cois et al. (2007)]{francois2007} Fran\c cois, P., Depagne, E., Hill, V., Spite, M. et al. 2007  \aap, 476,935 

\bibitem[Frebel (2010)]{frebel2010} Frebel , A 2010, Astronomische Nachrichten, 331, 474

\bibitem[Frischknecht et al. (2012)]{frischknecht2012} Frischknecht, U., Hirschi, R.,  Thielemann, F.-K. 2012, \aap, 538, L2

\bibitem[Gilmore et al. (2013)]{gilmore2013} Gilmore, G.,  Norris, J. E., Monaco, L. , Yong, D., Wyse, R. F. G.  Geisler, D. \apj, 763, 61


\bibitem[Goldoni et al. (2006)]{goldo2006} Goldoni, P., Royer, F., Fran\c cois, P., Horrobin, M., Blanc, G., Vernet, J., Modigliani, A., Larsen, J. 2006 SPIE 6269, 2

\bibitem[Gratton et al. (1994)]{gratton1994} Gratton, R.G., Sneden, C. 1994 /aap 287, 927

\bibitem[Greco et al.  (2008)]{greco2008} Greco, C., Dall'Ora, M., Clementini, G., Ripepi, V., Di Fabrizio, L., Kinemuchi, K., Marconi, M., Musella, I., Smith, H. A., Rodgers, C. T., Kuehn, C., Beers, T. C., Catelan, M., Pritzl, B. J. 2008, \apj 675,73

\bibitem[Grevesse \& Sauval (2000)]{grevesse2000}Grevesse, N. \& Sauval, A. J. 2000, Origin of Elements in the Solar System, ed. O. Manuel, 261

\bibitem[Gustafsson  et al. (1975)]{gustafsson1975} Gustafsson, B., Bell, R. A., Eriksson, K., \& Nordlund, A. 1975, \aap, ~42, 407

\bibitem[Gustafsson  et al. (2003)]{gustafsson2003} Gustafsson, B., Edvardsson, B., Eriksson, K., et al. 2003, in Stellar Atmosphere
Modeling, ed. I. Hubeny, D. Mihalas, \& K. Werner, ASP Conf. Ser., 288, 331

\bibitem[Gustafsson  et al. (2008)]{gustafsson2008} Gustafsson, B., Edvardsson, B., Eriksson, K., Jorgensen, U. G., Nordlund, A., Plez, B. 2008, \aap, 486, 951

\bibitem[Hill et al. (2002)]{hill2002} Hill, V., Plez, B., Cayrel, R. et al. 2002, \aap, 387, 560

\bibitem[Jordi et al. (2006)]{jordi2006}  Jordi, K., Grebel, E. K., Ammon, K. \aap, 460, 339

\bibitem[Kelson (2003)]{kelson2003} Kelson, D. 2003 \pasp,115, 688

\bibitem[Kirby et al.  (2010)]{kirby2010} 	Kirby, E. N., Guhathakurta, P., Simon, J. D., Geha, M. C., Rockosi, C. M., Sneden, C., Cohen, J. G., Sohn, S. T., Majewski, S. R., Siegel, M. 2010 \apjs,191, 352

\bibitem[Kirby et al. (2011)]{kirby2011} Kirby, E. N., Lanfranchi, G. A., Simon, J. D., Cohen, J. G., Guhathakurta, P. 2011 \apj, 727, 78

\bibitem[Kirby et al. (2008)]{kirby2008} Kirby, E. N., Simon, J. D., Geha, M., Guhathakurta, P. and Frebel A. 2008 \apj, 685, L43

\bibitem[Koch et al.  (2008)]{koch2008} Koch, A., McWilliam, A., Grebel, E. K., Zucker, D. B., Belokurov, V. 2008, \apj, 688, L13

\bibitem[Koch et al. (2009)]{koch2009}Koch, Andreas, Wilkinson, Mark I., Kleyna, Jan T., Irwin, Mike, Zucker, Daniel B., Belokurov, Vasily, Gilmore, Gerard F., Fellhauer, Michael, Evans, N. Wyn 2009, \apj,690, 453

\bibitem[Koch et al. (2013)]{koch2013}Koch, A., Feltzing, S., Ad\' en, D., Matteucci, F. 2013, \aap, 554, 5

\bibitem[Koch \& Rich (2014)]{koch2014}Koch, Andreas, Rich, R. Michael 2014, \apj,  794, 89

\bibitem[Kuehn et al.  (2008)]{kuehn2008}Kuehn, C., Kinemuchi, K., Ripepi, V., Clementini, G., Dall'Ora, M.,
 Di Fabrizio, L., Rodgers, C. T., Greco, C., Marconi, M., Musella, I., Smith, H. A., Catelan, M., Beers, T.C., Pritzl, B. 2008  \apj 674, 81

\bibitem[Lemasle et al. (2008)]{lemasle2008} Lemasle, B., Fran\c cois, P., Piersimoni, A., Pedicelli, S., Bono, G., Laney, C. D., Primas, F., Romaniello, M. 2008 \aap 490, 613

\bibitem[Lodders et al. (2009)]{lodders2009} Lodders, L., Palme, H., Gail, H-P, LanB 4, 44 Abundances of the elements in the Solar
System. Trumper, J.E. (ed.), Landolt-Bornstein (Springer-Verlag, Berlin)

\bibitem[Ludwig et al. (2008)]{ludwig2008} Ludwig, H.-G., Bonifacio, P., Caffau, E., Behara, N. T., Gonzalez Hernandez, J. I., Sbordone, L. 2008  \physscr, 133

\bibitem[McWilliam and Preston (1995)]{mcWilliam1995} McWilliam, A., Preston, G. 1995  \aj, 106,2757
\bibitem[Martin et al. (2008)]{martin2008}Martin, Nicolas F.
Coleman, M. G., De Jong, J. T. a., Rix, H-W., Bell, E. F.,Sand, D. J., Hill, J. M., Thompson, D., Burwitz, V., Giallongo, E., Ragazzoni, R., Diolaiti, E., Gasparo, F., Grazian, A., Pedichini, F., Bechtold, J. 2008 \apj, 672, L13

\bibitem[Moore et al. (1999)]{moore1999}  Moore, B., Ghigna, S., Governato, F. et al, \apjl, 524, L19

\bibitem[Moretti et al. (2009)]{moretti2009} Moretti, M. I., Dall'Ora, M., Ripepi, V., Clementini, G., Di Fabrizio, L., Smith, H., De Lee, N., Kuehn, C., Catelan, M., Marconi, M.,Musella, I.,Beers, T.,Kinemuchi, K.  2009 \apj, 699, 125

\bibitem[Musella et al. (2012)]{musella2012} 	Musella, I., Ripepi, V., Marconi, M., Clementini, G., Dall'Ora, M.,
 Scowcroft, V., Moretti, M. I., Di Fabrizio, L., Greco, C., Coppola, G., Bersier, D., Catelan, M., Grado, A., Limatola, L., Smith, H. A., Kinemuchi, K.
2012, \apj 756, 121

\bibitem[Okamoto et al. (2012)]{okamoto2012} Okamoto, S., Arimoto, N., Yamada, Y., Onodera, M. 2012, \apj, 744, 96

\bibitem[Pignatari et al. (2008)]{pignatari2008} Pignatari, M., Gallino, R., Meynet, G.,  Hirschi, R., Herwig, F., Wiescher, M. 2008, \aap, 687, L95 


\bibitem[Plez et al. (1992)]{plez1992} Plez, B., Brett, J. M., \& Nordlund, A.. 1992, \aap, 256, 551


\bibitem[Ram\' irez \& Mel\' endez (2005)]{ramirez2005} Ram\' irez, Iv\' an, Mel\' endez, Jorge 2005 \apj, 626, 465

\bibitem[Ricotti \& Gnedin (2005)]{ricotti2005} Ricotti, M. \& Gnedin, N. Y. 2005 \apj, 629, 259

\bibitem[Roderick et al.(2015)]{rode2015} Roderick , T.A., Jerjen H., Mackey A. D. , Da Costa G.  arXiv : 1503.03896


\bibitem[Schlegel et al. (1998)]{schlegel1998} Schlegel, D. J., Finkbeiner, D. P., Davis, M. 1998 \apj, 500, 525

\bibitem[Simon et al.(2007)]{simon2007}Simon, J. D., Geha, M. 2007, \apj, 670, 313

\bibitem[Simon et al. (2010)]{simon2010}Simon, J. D., Frebel, A., McWilliam, A., Kirby, E. N., Thompson, I. B. 2010, \apj, 716, 446

\bibitem[Ural et al. (2010)]{ural2010} Ural, Ugur, Wilkinson, M. I., Koch, A., Gilmore, G., Beers, T. C., Belokurov, V., Evans, N. W,
 Grebel, E. K., Vidrih, S., Zucker, D. B. 2010 \mnras 402,1357

\bibitem[Van Dokkum (2001)]{vandok2001} Van Dokkum,  P.  2001 \pasp, 562, 35
 

\bibitem[Vargas et al. (2013)]{vargas2013} Vargas, L. C., Geha, M., Kirby, E. N., Simon, J. D. 2013, \apj, 767, 134

\bibitem [Vernet et al. (2011)]{vernet2011} Vernet, J., Dekker, H., D'Odorico, S 2011 \aap, 536, 105

\bibitem[Vincenzo et al. (2014)]{vincenzo2014} Vincenzo, F., Matteucci, F., Vattakunnel, S., Lanfranchi, G. A. 2014 MNRAS 441, 2815

\bibitem[Wanajo (2013)]{wanajo2013} Wanajo, S., 2013 , \apjl 770, L12 

\bibitem[Weisz et al. (2014)]{weisz2014} Weisz, Daniel R., Dolphin, Andrew E., Skillman, Evan D., Holtzman, Jon, Gilbert, Karoline M., Dalcanton, Julianne J., Williams, Benjamin F. 2014, \apj, 789, 148

\bibitem[Weisz et al. (2015)]{weisz2015}Weisz, D. R., Dolphin, A. E., Skillman, E. D., Holtzman, J., Gilbert, K. M., Dalcanton, J. J., Williams, B. F. \apj, 804,136

\bibitem[Walsh et al. (2007)]{walsh2007} Walsh, S.M., Jerjen, H., Willman, B. 2007, \apjl, 662, L83

\bibitem[Walsch et al. (2008)]{walsh2008} Walsh, S., Jerjen, H., Willman, B. 2008  	
	Galaxies in the Local Volume, Astrophysics and Space Science Proceedings, ISBN 978-1-4020-6932-1. Springer Netherlands, 2008, p. 191

\bibitem[Webster et al. (2015)]{webster2015} Webster, D.,  Bland-Hawthorn, J.,  Sutherland, . 2015 \apj 799, 21
	
\bibitem[ Zucker et al. (2006)]{zucker2006} Zucker, D B, Belokurov, V, Evans, N W, Wilkinson, M I, Irwin, M J, Sivarani, T, Hodgkin, S, Bramich, D M, Irwin, J M, Gilmore, G, Willman, B, Vidrih, S, Fellhauer, M, Hewett, P C, Beers, T C, Bell, E F, Grebel, E K, Schneider, D P, Newberg, H J, Wyse, R F G, Rockosi, C M, Yanny, B, Lupton, R	
 2006, \apj, 643, 103	
 
\end{thebibliography}
\end{document}